\documentclass[a4paper,journal]{IEEEtran}

\usepackage{cite}
\usepackage{amsmath,amssymb,amsfonts}
\usepackage{algorithmic}
\usepackage{graphicx}
\usepackage{subfigure}
\usepackage{textcomp}
\usepackage{xcolor}
\usepackage{url}
\usepackage{bbm}
\usepackage{bm}
\usepackage{booktabs}

\usepackage{algorithm}
\usepackage{array}
\usepackage{stfloats}
\usepackage{verbatim}

\usepackage{hyperref}

\begin{document}
\title{Time-Frequency-Space Transmit Design and Receiver Processing with Dynamic Subarray for Terahertz Integrated Sensing and Communication}

\author{Yongzhi~Wu,~\IEEEmembership{Graduate~Student~Member,~IEEE,}
        Chong~Han,~\IEEEmembership{Senior Member,~IEEE,} and Meixia~Tao,~\IEEEmembership{Fellow,~IEEE}
\thanks{
This paper was presented in part at IEEE SPAWC, September 2023~\cite{wu2023signal}.

Yongzhi~Wu is with the Terahertz Wireless Communications (TWC) Laboratory, Shanghai Jiao Tong University, Shanghai, China (Email:~yongzhi.wu@sjtu.edu.cn). 

Chong~Han is with the Terahertz Wireless Communications (TWC) Laboratory, Department of Electronic Engineering and Cooperative Medianet Innovation Center (CMIC), Shanghai Jiao Tong University, Shanghai, China (Email:~chong.han@sjtu.edu.cn). 

Meixia Tao is with the Department of Electronic Engineering, Shanghai Jiao Tong University, Shanghai, China (Email:~mxtao@sjtu.edu.cn).}
}

\maketitle
\thispagestyle{empty}
\pagestyle{empty}


\begin{abstract}
Terahertz (THz) integrated sensing and communication (ISAC) enables simultaneous data transmission with Terabit-per-second (Tbps) rate and millimeter-level accurate sensing. To realize such a blueprint, ultra-massive antenna arrays with directional beamforming are used to compensate for severe path loss in the THz band. 
In this paper, the time-frequency-space transmit design is investigated for THz ISAC to generate time-varying scanning sensing beams and stable communication beams. Specifically, with the dynamic array-of-subarray (DAoSA) hybrid beamforming architecture and multi-carrier modulation, two ISAC hybrid precoding algorithms are proposed, namely, a vectorization (VEC) based algorithm that outperforms existing ISAC hybrid precoding methods and a low-complexity sensing codebook assisted (SCA) approach. Meanwhile, coupled with the transmit design, parameter estimation algorithms are proposed to realize high-accuracy sensing, including a wideband DAoSA MUSIC method for angle estimation and a sum-DFT-GSS approach for range and velocity estimation. Numerical results indicate that the proposed algorithms can realize centi-degree-level angle estimation accuracy and millimeter-level range estimation accuracy, which are one or two orders of magnitudes better than the methods in the millimeter-wave band. In addition, to overcome the cyclic prefix limitation and Doppler effects, an inter-symbol interference- and inter-carrier interference-tackled sensing algorithm is developed to refine sensing capabilities for THz ISAC.
\end{abstract}


\section{Introduction}
%
%
%
%
\subsection{Background and Motivations}

\IEEEPARstart{T}{o} address the rapidly growing demand for wireless data rates and the emergence of new application scenarios, the communication community is seeking new spectrum opportunities as well as new functionalities for sixth-generation (6G) and beyond wireless networks~\cite{Tong20216G}. Following the former trend of moving up to higher frequencies, the Terahertz (THz) band is viewed as one of the key technologies to enable enormous potential in the next-generation advanced transceiver (NGAT)~\cite{Akyildiz2022THz}. Another promising usage scenario supported by the NGAT is integrated sensing and communication (ISAC), which can endow wireless networks with sensing capabilities to realize the mapping of the physical world to the digital world~\cite{liu2022ISAC}. 
Leveraging the ultra-broad bandwidth and the ultra-massive antenna arrays in the THz band, the integration of these two technologies, i.e., \textit{Terahertz integrated sensing and communication (THz ISAC)}~\cite{han2023thz-isac}, can achieve ultra-accurate sensing and Terabit-per-second data rates simultaneously.

Despite the promising vision of THz ISAC, critical challenges arise when designing THz ISAC transmit signals. First, there exists severe path loss in the THz band, which includes free path loss, reflection, and scattering losses. These losses strictly limit the maximum sensing and communication distance, and degrade sensing accuracy and data rate.
Second, with the power constraints, to compensate for such severe path loss, ultra-massive multiple-input multiple-output (UM-MIMO) antenna arrays with beamforming are used to generate highly directional beams~\cite{han2021hybrid}. Thus, energy-efficient and low-complexity beamforming algorithms need to be developed.
Third, the generation of directional beams restricts the angular coverage of sensing. In general, communication prefers stable beams toward users to enable tractable data detection, while sensing requires sweeping beams to scan possible targets in the surrounding environment~\cite{zhang2022jcrs}. To realize entire-space sensing with directional beams, effective and efficient narrowbeam management schemes, including transmit design in the time-frequency domain and beamforming design in the spatial domain are demanded to realize simultaneous sensing and communication for THz ISAC systems.

Meanwhile, the receive processing encounters significant challenges, especially for sensing parameter estimation algorithms in THz UM-MIMO systems, which are affected by the beamforming architectures and peculiarities of THz channels. First, the sensing algorithm for range and velocity estimation needs to be redesigned, since an additional dimension (namely, spatial domain) is introduced in the received signal model when using the ultra-large dimensional antenna arrays in the THz band.
Second, with high channel sparsity due to strong power loss of non-line-of-sight (NLoS) paths, the delay spread of the THz communication channel is reduced~\cite{han2022channel}. In this case, to utilize broad bandwidth with a fixed subcarrier number, we can increase subcarrier spacing, which is inversely proportional to the symbol duration. Thus, the symbol duration and cyclic prefix (CP) length are reduced in classical multi-carrier communication systems, such as orthogonal frequency-division multiplexing (OFDM). Nevertheless, the round-trip delay of sensing targets should be smaller than the CP duration with classical OFDM sensing algorithms~\cite{sturm2011waveform}. For communication waveforms with reduced CP, there might exist inter-symbol interference (ISI) effects on the received sensing signal, which cause existing sensing methods inapplicable.
Third, as the Doppler shifts are proportional to the carrier frequency, the Doppler effects become even stricter in the THz band. If maintaining current waveform numerology of 5G wireless systems, Doppler effects in the presence of high-mobility targets may cause inter-carrier interference (ICI) effects and severely degrade sensing capabilities. Thus, to tackle these challenges, signal processing design in terms of sensing algorithms is vital to realize high-accuracy sensing, while data recovery has been well investigated~\cite{sarieddeen2021overview}.

\subsection{Related Works}

\subsubsection{Waveform Design}
By jointly designing the ISAC transmit signal, sensing and communication can share the hardware and signal processing modules. From the perspective of the time-frequency domain, various ISAC waveforms have been investigated in the literature. As adopted in 4G and 5G standards, CP-OFDM is a promising candidate for ISAC although being a communication-centric design~\cite{berger2010ofdm, sturm2011waveform, johnston2022ofdm, keskin2021jrc, zhang2019jcrs, wukai2022isac}. Since an OFDM waveform suffers from a high peak-to-average ratio (PAPR) issue, especially in uplink transmission, some single-carrier waveforms, such as discrete Fourier transform (DFT) spread OFDM (DFT-s-OFDM), are investigated for THz ISAC systems, due to their low PAPR compared to OFDM~\cite{wu2023dftsofdm}. Recently, orthogonal time frequency space (OTFS) has been studied in ISAC applications~\cite{lorenzo2020otfs, dehkordi2023otfs, wukai2023otfs}, thanks to its advantages under doubly-selective channels in high-mobility scenarios. Furthermore, a DFT spread OTFS (DFT-s-OTFS) waveform is proposed in~\cite{wu2023dftsotfs} to reduce the PAPR of OTFS for THz ISAC. However, the high complexity of data detection for MIMO-OTFS constitutes a serious problem. Despite the PAPR issue, OFDM is still a potential waveform in the THz band, since it has good compatibility with UM-MIMO and enables flexible time-frequency domain resource allocation among multiple users~\cite{han2023thz-isac}. Thus, wideband UM-MIMO systems with multi-carrier modulations are investigated for THz communications in many recent works, including beamforming design~\cite{yuan2018beamforming, yuan2020beamforming, yan2022beamforming, gao2021beamforming}, channel estimation~\cite{dovelos2021mimo}, multiple access~\cite{zhai2021ofdma}, carrier aggregation~\cite{samara2023ofdm}. Nevertheless, there is a lack of research on THz ISAC in this regard, especially focusing on the transmit design and sensing algorithms in the time-frequency-space domain.

\subsubsection{Beamforming Design}
Pertaining to MIMO-OFDM systems, with conventional fully-digital and analog beamforming architectures, multi-target estimation can be realized by utilizing opportunistic sensing~\cite{keskin2021jrc} and multibeam optimization~\cite{luo2019jcrs}. 
Nevertheless, the fully-digital structure exhibits high hardware complexity and power consumption for THz ISAC systems with large-dimensional antenna arrays, while the analog beamforming architecture can only support one data stream with limited spatial multiplexing gain~\cite{han2021hybrid}.
As a combined approach, hybrid beamforming can realize comparable data rates with the fully-digital structure and exhibits less hardware complexity. Based on the full-connected (FC) hybrid beamforming architecture, authors in \cite{cheng2021dfrc} propose a consensus-ADMM approach to design the analog and digital beamformers by jointly optimizing the spectral efficiency (SE) and spatial spectrum matching error of sensing. With the array-of-subarray (AoSA) structure, which further reduces the number of phase shifters and power consumption at the cost of sacrificing data rate, the ISAC hybrid beamformers can be designed by optimizing the Cram\'er-Rao bound~\cite{wang2022isac} or minimizing the weighted Euclidean distance between the hybrid precoding matrix and the fully digital beamforming matrix~\cite{liu2019jsc}. To balance SE and power consumption, a dynamic array-of-subarray (DAoSA) hybrid precoding architecture is proposed in~\cite{yan2020DAoSA}, while the ISAC hybrid precoding design with dynamic subarray has not been investigated yet.
In addition, most of the aforementioned works design beamformers with some prior knowledge of target angles~\cite{keskin2021jrc, cheng2021dfrc, wang2022isac, liu2019jsc, Elbir2021jrc}, which is acceptable in target tracking scenarios but not available in general target estimation, i.e., target discovery mode. Thus, beam scanning-based sensing to discover targets with narrow beams in the THz band is still a significant issue to be addressed.

\begin{figure*}
    \centering
    \includegraphics[width=0.86\textwidth]{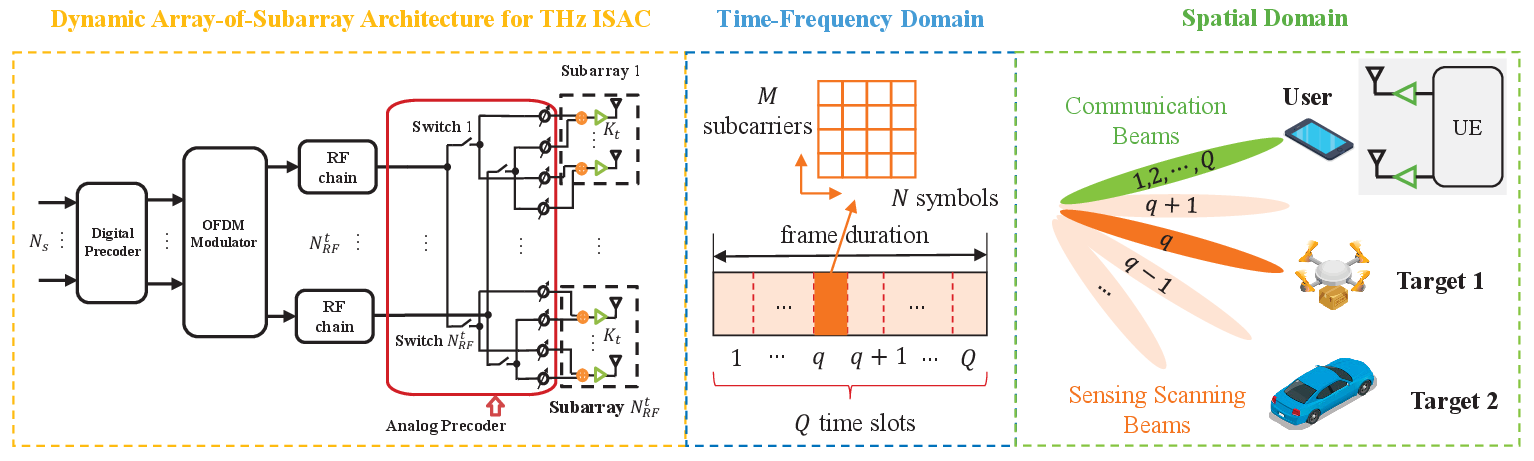}
    \caption{Based on the DAoSA architecture, the proposed THz ISAC system generates scanning beams toward the $q$\textsuperscript{th} sensing direction and stable beams toward the communication user at the $q$\textsuperscript{th} time slot.}
    \label{fig:system_model}
\end{figure*}

\subsection{Contributions and Paper Structure}

The contributions of this work are summarized as follows:
\begin{itemize}
    \item We present a time-frequency-space transmit design framework for THz ISAC systems by considering a dynamic subarray hybrid beamforming architecture and multi-carrier waveform. In this framework, we develop a vectorization (VEC) based and a sensing codebook-assisted (SCA) ISAC hybrid precoding algorithms for the DAoSA structure. Our proposed ISAC hybrid precoding algorithms can realize sensing of the entire angular directions and data transmission by generating scanning sensing beams at different time slots and stable communication beams toward the user. Meanwhile, the proposed VEC algorithm outperforms existing ISAC hybrid precoding methods, and the SCA approach reduces the computational complexity.
    \item Based on the time-frequency-space domain transmit signal design, we propose parameter estimation algorithms at the sensing receiver, including a wideband DAoSA MUSIC (W-DAoSA-MUSIC) algorithm for angle estimation, and a sum-DFT and golden section search (S-DFT-GSS) method for range and velocity estimation. Simulation results indicate that the sensing accuracy with the proposed sensing algorithms can achieve centi-degree-level for angle estimation, millimeter-level for range estimation, and decimeter-per-second-level for velocity estimation.
    \item We further propose an ISI- and ICI-tackled sensing algorithm to overcome the CP limitation on the maximum sensing distance and estimation error caused by high-mobility targets. While the ICI is studied in~\cite{keskin2021jrc}, the ISI effects have not been considered in the literature. Compared to the ISI-unaware estimation, the ISI-tackled sensing algorithm can accurately estimate the target with a round-trip delay larger than the CP duration. In contrast with ICI-unaware estimation, the ICI-tackled algorithm can overcome the masking problem of weak targets caused by the side lobes of the strong target in the presence of ICI effects.
\end{itemize}

The structure of the remainder of this paper is organized as follows. The system framework with the time-frequency-space transmit design for THz ISAC is presented in Sec.~\ref{sec:system}. The ISAC hybrid precoding algorithms are elaborated in Sec.~\ref{sec:hybrid_precoding}. The sensing estimation algorithm design with the DAoSA architecture and multi-carrier modulation is proposed in Sec.~\ref{sec:sensing_algorithm}. The ISI- and ICI-tackled sensing algorithm for THz ISAC is developed in Sec.~\ref{sec:isi_ici}. Sec.~\ref{sec:simulation} illustrates extensive simulation results. Finally, the paper is concluded in Sec.~\ref{sec:conclusion}. 

\textbf{Notations:} $\mathbb{C}$ denotes the set of complex numbers; $\mathbf{A}(i, j)$ is the entry on the $i$\textsuperscript{th} row and $j$\textsuperscript{th} column of $\mathbf{A}$; $\mathbb{E}\{\cdot\}$ defines the expectation operation; The superscripts $(\cdot)^T$ and $(\cdot)^H$ stand for the transpose and Hermitian transpose operations; The notations $\otimes$ and $\odot$ refer to the Kronecker product and Hadamard Product, respectively; $\text{det}(\cdot)$ and $\|\cdot\|_F$ denote the determinant and Frobenius norm of a matrix; $(\cdot)^\dagger$ indicates the Moore-Penrose pseudo inverse; $\text{vec}(\cdot)$ represents the vectorization operation.

\section{System Framework}\label{sec:system}

As shown in Fig.~\ref{fig:system_model}, we propose a THz ISAC system framework based on a wideband UM-MIMO architecture, in which the ISAC transceiver simultaneously senses potential targets in the surrounding spatial environment and sends information symbols to one communication receiver (without loss of generality) via the designed transmit signal in the time-frequency-space domain. Specifically, in the time-frequency domain, the data signal is modulated with orthogonal frequency-division multiplexing (OFDM) and spread across $M$ subcarriers. In the spatial domain, the data streams at each subcarrier are precoded through a digital precoder $\mathbf{F}_\text{BB} \in \mathbb{C}^{N_\text{RF}^t\times N_s}$ and an analog precoder $\mathbf{F}_\text{RF}\in \mathbb{C}^{N_t \times N_\text{RF}^t}$, where $N_s$ denotes the number of data streams and $N_\text{RF}^t$ refers to the number of transmit RF chains, with $N_s \leqslant N_\text{RF}^t \ll N_t$. 

As for the transceiver structure, the ISAC transceiver is equipped with an $N_t$-element transmit uniform planar array (UPA) to transmit the ISAC waveform and an $N_r$-element receive UPA to perform sensing echo processing. The communication receiver has an $N_r$-element UPA to accomplish signal reception and data detection. The transmit antenna arrays adopt a DAoSA hybrid beamforming structure~\cite{yan2020DAoSA}. With the DAoSA structure, the transmit antennas are divided into $N_\text{RF}^t$ subarrays and each RF chain connects to each subarray with $K_t = N_t / N_\text{RF}^t$ elements through a switch. Similarly, the received signal is combined through the analog combiner and the digital combiner with $N_\text{RF}^r$ RF chains, and each receiver subarray contains $K_r = N_r / N_\text{RF}^r$ elements.

\subsection{Time-Frequency-Space Transmit Design}

At the transmitter side, the ISAC system maps the transmitted bit streams to a large amount of data frames. A data frame is divided into $Q$ time slots, each of which contains $M \times N$ data symbols, where $M$ and $N$ stand for the numbers of subcarriers and symbols during a time slot. In the multi-carrier hybrid beamforming architecture, at the $q$\textsuperscript{th} time slot, the data symbols $\mathbf{s}_q[m, n] \in \mathbb{C}^{N_s\times 1}, q = 1, 2, \cdots, Q, m = 0, 1, \cdots, M - 1, n = 0, 1, \cdots, N - 1$, which are generated from $N_s$ data streams with $\mathbb{E}\{\mathbf{s}_q[m, n] \mathbf{s}^H_q[m, n]\} = \frac{1}{N_s} \mathbf{I}_{N_s}$, are first precoded by a digital beamformer $\mathbf{F}_{\text{BB}, q}[m]$ and mapped to the $m$\textsuperscript{th} subcarrier in the frequency domain, $\mathbf{x}_q[m, n] = \mathbf{F}_{\text{BB}, q}[m] \mathbf{s}_q[m, n]$. Then, we perform the inverse discrete Fourier transform (IDFT) to transform the frequency-domain data blocks to the time-domain signal and add one cyclic prefix (CP) for each symbol before conducting up-conversion and analog beamforming $\mathbf{F}_{\text{RF}, q}\in \mathbb{C}^{N_t\times N_\text{RF}^t}$. 

At the $q$\textsuperscript{th} time slot, the proposed THz ISAC system with the time-frequency-space three-dimensional transmit design generates scanning beams toward the $q$\textsuperscript{th} sensing direction and stable beams toward the communication user.
Note that all subcarriers share the same analog precoder while the digital precoder is performed for each subcarrier~\cite{yu2016altmin}.
For the $n$\textsuperscript{th} symbol during the $q$\textsuperscript{th} time slot, the transmit time-domain signal can be expressed as,
\begin{equation}
    \mathbf{\tilde{x}}_{q, n} (t) = \sum_{m=0}^{M-1} \mathbf{F}_{\text{RF}, q} \mathbf{F}_{\text{BB}, q}[m] \mathbf{s}_q[m, n] e^{j2\pi m\Delta f t},
\end{equation}
where $t$ denotes the time instant and $\Delta f$ refers to the subcarrier spacing. Then, the symbol duration $T$ equals to $\frac{1}{\Delta f}$ and the total symbol duration is expressed as $T_o = T + T_\text{cp}$ with the CP duration of $T_\text{cp} = \frac{M_\text{cp}}{M} T$, where $M_\text{cp}$ is the CP size. Thus, the duration of a time slot is $T_s = N T_o$ and the frame duration can be expressed as $T_f = Q T_s$. To generate stable beams towards the communication user and scanning beams for searching sensing targets, the transmit beamformers are fixed during a time slot and vary at different time slots.   

In this work, we consider a DAoSA hybrid beamforming architecture~\cite{yan2020DAoSA}, in which the connections between RF chains and subarrays can be intelligently adjusted through a network of switches. The analog precoding matrix $\mathbf{F}_{\text{RF}, q}$ can be written as,
\begin{equation}
    \mathbf{F}_{\text{RF}, q} = \mathbf{F}_{\text{P}, q} \odot \mathbf{P}_{\text{S}},
\end{equation}
where $\mathbf{F}_{\text{P}, q} \in \mathbb{C}^{N_t\times N_\text{RF}^t}$ denotes the phase shifter network matrix and $\mathbf{P}_{\text{S}} \in \{0, 1\}^{N_t \times N_\text{RF}^t}$ describes the binary switch network matrix, which can be expressed as
\begin{equation}
    \mathbf{P}_{\text{S}}=\left[\begin{array}{cccc}
\mathbf{p}_{1,1} & \mathbf{p}_{1,2} & \ldots & \mathbf{p}_{1, N_\text{RF}^t} \\
\mathbf{p}_{2,1} & \mathbf{p}_{2,2} & \ldots & \mathbf{p}_{2, N_\text{RF}^t} \\
\vdots & \vdots & \ddots & \vdots \\
\mathbf{p}_{N_\text{RF}^t, 1} & \mathbf{p}_{N_\text{RF}^t, 2} & \ldots & \mathbf{p}_{N_\text{RF}^t, N_\text{RF}^t}
\end{array}\right],
\end{equation}
where $\mathbf{p}_{i, j}$ stands for the status of the switch between the $i$\textsuperscript{th} subarray and the $j$\textsuperscript{th} RF chain. If this switch is closed, $\mathbf{p}_{i, j} = \mathbf{1}_{K_t}$ is an all-one vector. Conversely,  $\mathbf{p}_{i, j} = \mathbf{0}_{K_t}$ is a zero vector. The phase shifter network matrix $\mathbf{F}_{\text{P}, q}$ satisfies a
constant modulus constraint, i.e., the modulus of its elements is 1. Then, the analog precoding matrix $\mathbf{F}_{\text{RF}, q}$ is given by
\begin{equation}
        \mathbf{F}_{\text{RF}, q}=\left[\begin{array}{cccc}
\mathbf{f}_{1,1} & \mathbf{f}_{1,2} & \ldots & \mathbf{f}_{1, N_\text{RF}^t} \\
\mathbf{f}_{2,1} & \mathbf{f}_{2,2} & \ldots & \mathbf{f}_{2, N_\text{RF}^t} \\
\vdots & \vdots & \ddots & \vdots \\
\mathbf{f}_{N_\text{RF}^t, 1} & \mathbf{f}_{N_\text{RF}^t, 2} & \ldots & \mathbf{f}_{N_\text{RF}^t, N_\text{RF}^t}
\end{array}\right],
\end{equation}
where $\mathbf{f}_{i, j}\in \mathbb{C}^{K_t \times 1}$ represents the joint precoding vector of the switch and the phase shifters between the $i$\textsuperscript{th} subarray and the $j$\textsuperscript{th} RF chain. When this switch is closed, $\mathbf{f}_{i, j}$ should satisfy the unit modulus constraint. When the switch is open, $\mathbf{f}_{i, j}$ is a zero vector. We denote the feasible set of the analog precoder $\mathbf{F}_{\text{RF}, q}$ as $\mathcal{F}$. Moreover, the normalized transmit power constraint is expressed as, $\|\mathbf{F}_{\text{RF}, q} \mathbf{F}_{\text{BB}, q}[m]\|_F^2 = N_s$.

\subsection{Communication Model}

With multi-carrier transmission, the communication received signal of the $m$\textsuperscript{th} subcarrier and the $n$\textsuperscript{th} symbol at $q$\textsuperscript{th} time slot after the decoding process is expressed as
\begin{equation}
\begin{split}
        \mathbf{r}_q[m, n] =& \sqrt{\rho} \mathbf{C}_{\text{BB}}^H[m] \mathbf{C}_{\text{RF}}^H \mathbf{H}_c[m] \mathbf{F}_{\text{RF}, q} \mathbf{F}_{\text{BB}, q}[m] \mathbf{s}_q[m, n] \\
        &+ \mathbf{C}_{\text{BB}}^H[m] \mathbf{C}_{\text{RF}}^H \mathbf{n}_q[m, n],
\end{split}
\end{equation}
where $\rho$ describes the average received power, $\mathbf{C}_{\text{BB}}[m]\in \mathbb{C}^{N_\text{RF}^r\times N_s}$ is the digital combining matrix, $\mathbf{C}_{\text{RF}}\in \mathbb{C}^{N_r \times N_\text{RF}^r}$ is the analog combining matrix, and $\mathbf{n}_q[m, n]$ refers to the additive white Gaussian noise with independent and identically distribution $\mathcal{CN}(0, \sigma_n^2)$. In the THz band, the channel is sparse and dominated by the line-of-sight (LoS) path and several reflected rays. Thus, as a benchmark, the multi-path channel model based on ray-tracing methods of the channel matrix $\mathbf{H}_c[m]$ at the $m$\textsuperscript{th} subcarrier can be given by \cite{yan2020DAoSA, Elbir2021jrc, yuan2020beamforming},
\begin{equation}
\begin{split}
        \mathbf{H}_c[m] =& \gamma \alpha_\text{L}[m] \mathbf{a}_r\left(\theta_\text{L}^r, \phi_\text{L}^r\right) \mathbf{a}_t^H\left(\theta_\text{L}^t, \phi_\text{L}^t\right) \\
        & + \gamma \sum_{l=1}^{L_\text{N}} \alpha_{\text{N}, l}[m] \mathbf{a}_r\left(\theta_{\text{N}, l}^r, \phi_{\text{N}, l}^r\right) \mathbf{a}_t^H\left(\theta_{\text{N}, l}^t, \phi_{\text{N}, l}^t\right),
\end{split}
\end{equation}
where $\gamma = \sqrt{\frac{N_t N_r}{L_\text{N} + 1}}$ and $L_\text{N}$ represents the number of non-line-of-sight (NLoS) paths. Moreover, $\alpha_\text{L}[m]$ and $\alpha_{\text{N}, l}[m]$ denote the channel gain of the LoS path and $l$\textsuperscript{th} NLoS path at $m$\textsuperscript{th} subcarrier, respectively. In addition, $\theta^r$($\theta^t$) and $\phi^r$($\phi^t$) refer to the azimuth and elevation angles of arrival/departure (AoAs/AoDs). In the case of a UPA in the $yz$-plane with $W$ and $L$ elements on the $y$ and $z$ axes respectively, the array response vector can be expressed by,
\begin{equation}
\begin{split}
    \mathbf{a}(\theta, \phi) &= \mathbf{a}_z(\phi) \otimes \mathbf{a}_y(\theta, \phi),
\end{split}
\end{equation}
where
\begin{equation}
    \mathbf{a}_y(\theta, \phi) = \frac{1}{\sqrt{W}} [1, \cdots, e^{j\pi (W - 1) \sin(\theta) \sin(\phi)}]^T,
\end{equation}
\begin{equation}
    \mathbf{a}_z(\phi) = \frac{1}{\sqrt{L}} [1, \cdots, e^{j\pi (L - 1) \cos(\phi)}]^T,
\end{equation}
and $\theta$ stands for the azimuth angle, and $\phi$ refers to the elevation angle. When the ratio between the used bandwidth and central frequency is small, the beam squint effect is limited. With the 1024-element UPA at 0.3 THz, the Rayleigh distance is about 1~m, so the near-field effect can be ignored in this paper.

For THz communications, we need to design hybrid precoders to maximize spectral efficiency. The achievable spectral efficiency can be expressed as~\cite{yan2020DAoSA}
\begin{equation}
    \begin{split}
        R_q =& \frac{1}{M} \sum_{m=0}^{M-1}\log \det\bigg(\mathbf{I}_{N_s} + \frac{\rho}{N_s} \mathbf{R}_n^{-1} \mathbf{C}_{\text{BB}}^H[m] \mathbf{C}_{\text{RF}}^H \mathbf{H}_c[m] \\
        &\times \mathbf{F}_{\text{RF}, q} \mathbf{F}_{\text{BB}, q}[m] \mathbf{F}_{\text{BB}, q}^H[m]  \mathbf{F}_{\text{RF}, q}^H \mathbf{H}_c^H[m] \mathbf{C}_{\text{RF}} \mathbf{C}_{\text{BB}}[m]\bigg),
    \end{split}
\end{equation}
where $\mathbf{R}_n = \sigma_n^2 \mathbf{C}_{\text{BB}}^H[m] \mathbf{C}_{\text{RF}}^H \mathbf{C}_{\text{RF}} \mathbf{C}_{\text{BB}}[m]$ is a noise covariance matrix. The optimization problem of maximizing $R_q$ at the transmitter side is equivalent to minimizing the Euclidean distance between the optimal fully digital precoder $\mathbf{F}_c[m]$ and the hybrid precoder as $\frac{1}{M}\sum_{m=0}^{M-1}\|\mathbf{F}_c[m] - \mathbf{F}_{\text{RF}, q}\mathbf{F}_{\text{BB}, q}[m]\|_F^2$. Generally, the channel state information (CSI) can be known at both transmitter and receiver by utilizing channel estimation~\cite{chen2022mimo} and is assumed to be time-invariant during a frame duration. Then, from the singular value decomposition (SVD) of the channel $\mathbf{H}_c[m]$, the unconstrained optimal precoder $\mathbf{F}_c[m]$ and decoder $\mathbf{C}_c[m]$ are comprised of the first $N_s$ columns of the right and the left singular value matrices.

\subsection{Sensing Model}\label{sec:sensing_model}

In the THz band, directional beams are used to compensate for severe path loss and improve received sensing signal power, which limits the angular range of sensing targets. To realize entire-space sensing, we design a codebook-based beam-scanning scheme for THz sensing.
For the azimuth angle, the whole sensing angular domain is divided into $Q$ scanning directions, $\mathbf{\omega} = [\omega_1, \omega_2, \cdots, \omega_Q]^T$, each of which corresponds to a time slot. We can set $Q = W$ and design the sensing beamforming vector as the $q$\textsuperscript{th} column from a discrete Fourier transform (DFT) codebook, by which the transmitter can generate $W$ orthogonal beamforming vectors and steer signals towards $W$ independent sensing directions. Thus, the sensing codebook can be written as,
\begin{equation}
    \mathbf{A} = \mathbf{a}_z(\phi) \otimes [\mathbf{a}_{y,1}(\omega_1, \phi), \cdots, \mathbf{a}_{y, W}(\omega_Q, \phi)]
\end{equation}
where
\begin{equation}
    \mathbf{a}_{y, q}(\omega_q, \phi) = \frac{1}{\sqrt{W}} [1, \cdots, e^{j\pi (W - 1)\sin(\omega_q)\sin(\phi)}]^T,
\end{equation}
and $\sin(\omega_q) = -1 + \frac{1}{W} + (q -1) \frac{2}{W}$ for $q = 1, 2, \cdots, W$. In this case, the sensing angular window $\Omega_q$ at the $q$\textsuperscript{th} time slot contains angles from $\arcsin(-1+(q-1)\frac{2}{W})$ to $\arcsin(-1+q\frac{2}{W})$.

At the sensing receiver, the frequency domain received signal of the $m$\textsuperscript{th} subcarrier and the $n$\textsuperscript{th} symbol at $q$\textsuperscript{th} time slot is denoted as $\mathbf{y}_q[m, n]\in\mathbb{C}^{N^r_\text{RF}\times 1}$, which is given by
\begin{equation}
\begin{split}
        \mathbf{y}_q[m, n] =& \mathbf{W}_{\text{RF}, q}^H \mathbf{H}_{s}[m, n] \mathbf{F}_{\text{RF}, q} \mathbf{F}_{\text{BB}, q}[m] \mathbf{s}_q[m, n] \\
        &+ \mathbf{W}_{\text{RF}, q}^H \mathbf{e}_q[m, n]
\end{split}
\end{equation}
where $\mathbf{W}_{\text{RF}, q}\in \mathbb{C}^{N_r\times N_\text{RF}^r }$ denotes the combing matrix at the sensing receiver and $\mathbf{e}_q[m, n]$ represents the AWGN vector.
At the ISAC transceiver side, the sensing receiver is collocated with the transmitter. Based on the OFDM radar sensing channel~\cite{sturm2011waveform, wu2023dftsofdm} and MIMO channel models~\cite{yan2020DAoSA, yuan2020beamforming}, the sensing channel matrix $\mathbf{H}_s[m, n]$ is expressed as,
\begin{equation}
\begin{split}
        \mathbf{H}_s[m, n] =& \sqrt{\frac{N_t N_r}{P}}\sum_{p=1}^P h_{p} e^{-j2\pi m \Delta f \tau_p} e^{j2\pi ((q - 1) T_s + n T_o) \nu_p} \\
        &\times \mathbf{a}_r(\theta_p, \phi_p) \mathbf{a}_t^T(\theta_p, \phi_p),
\end{split}
\end{equation}
where $P$ stands for the number of sensing targets, each of which corresponds to one back-reflected path with complex channel coefficient $h_p$. For the $p$\textsuperscript{th} target, the delay $\tau_p$ and the Doppler shift $\nu_p$ are calculated by $\tau_p = \frac{2 r_p}{c_0} (\tau_p \leqslant T_\text{cp})$ and $\nu_p = \frac{2 f_c v_p}{c_0} (\nu_p \ll \Delta f)$, where $r_p$ and $v_p$ refer to the range and relative velocity of the $p$ targets, respectively. $c_0$ denotes the speed of light and $f_c$ describes the carrier frequency. Moreover, $\theta_p$ and $\phi_p$ represent the azimuth and elevation angle-of-arrival of the $p$\textsuperscript{th} target.

Beamforming design for sensing aims at achieving the highest beamforming gain towards the sensing direction. Thus, at the $q$\textsuperscript{th} time slot, the optimal sensing precoder $\mathbf{F}_{s, q}\in\mathbb{C}^{N_t \times N_s}$ can be generated from the $q$\textsuperscript{th} column of the sensing codebook, namely, $\mathbf{F}_{s, q} = \frac{1}{\sqrt{N_t}}\mathbf{A}(:, q) \mathbf{1}_{N_s}^T$ with a normalized factor of $\frac{1}{\sqrt{N_t}}$. Then, we need to minimize the Euclidean distance, $\frac{1}{M}\sum_{m=0}^{M-1}\|\mathbf{F}_{s, q} - \mathbf{F}_{\text{RF}, q}\mathbf{F}_{\text{BB}, q}[m]\|_F^2$. At the sensing receiver side, $\mathbf{W}_{\text{RF}, q}$ is fixed during a time slot and the receive sensing beams point to $N_\text{RF}^r$ random directions within $\Omega_q$ at the $q$\textsuperscript{th} time slot.

\subsection{Problem Formulation}

At the THz ISAC transmitter, we need to design the analog and digital beamformers to simultaneously realize a communication link with ultra-fast data rates and provide a desired beampattern for high-accuracy sensing of surrounding targets.
Different from the conventional hybrid precoding design problem for communication, the optimal ISAC hybrid precoders should be sufficiently ``close" to the time-invariant and frequency-dependent optimal communication precoder and the time-varying and frequency-independent optimal sensing precoder at the same time.
Based on the above models and analysis, we can formulate the following multi-objective optimization problem,
\begin{equation}\label{eq:moop}
    \begin{split}
        \min_{\mathbf{F}_{\text{RF}, q}, \mathbf{F}_{\text{BB}, q}[m]} & \bigg\{\frac{1}{M}\sum_{m=0}^{M-1}\|\mathbf{F}_c[m] - \mathbf{F}_{\text{RF}, q}\mathbf{F}_{\text{BB}, q}[m]\|_F^2, \\
        & \frac{1}{M}\sum_{m=0}^{M-1}\|\mathbf{F}_{s, q} - \mathbf{F}_{\text{RF}, q}\mathbf{F}_{\text{BB}, q}[m]\|_F^2 ] \bigg\} \\
        \text{s.t. \quad } & \mathbf{F}_{\text{RF}, q} \in \mathcal{F},
        \|\mathbf{F}_{\text{RF}, q} \mathbf{F}_{\text{BB}, q}[m]\|_F^2 = N_s,\\
        & m = 0, 1, \cdots, M - 1,
    \end{split}
\end{equation}
for $q = 1, 2, \cdots, Q$.
Since this problem has multiple objective functions and the constraints are non-convex, it is rather difficult to obtain the global optimal solution. In the next section, we propose two algorithms for the THz ISAC hybrid precoding optimization problem to yield near-optimal solutions.

\section{Hybrid Precoding Design for THz ISAC}\label{sec:hybrid_precoding}

For the multi-objective ISAC hybrid precoding problem, we can introduce a weighting factor $\eta~(0 \leq \eta \leq 1)$, which provides the tradeoff between sensing and communication. Then, the hybrid precoding problem \eqref{eq:moop} can be formulated as,
\begin{equation}\label{eq:weighted_sum}
    \begin{split}
        \min_{\mathbf{F}_{\text{RF}, q}, \mathbf{F}_{\text{BB}, q}[m]} & \frac{1}{M}\sum_{m=0}^{M-1} \bigg(\eta \|\mathbf{F}_c[m] - \mathbf{F}_{\text{RF}, q}\mathbf{F}_{\text{BB}, q}[m]\|_F^2 + \\
        & (1 - \eta) \|\mathbf{F}_{s, q} - \mathbf{F}_{\text{RF}, q}\mathbf{F}_{\text{BB}, q}[m]\|_F^2 \bigg) \\
        \text{s.t. \quad } & \mathbf{F}_{\text{RF}, q} \in \mathcal{F},
        \|\mathbf{F}_{\text{RF}, q} \mathbf{F}_{\text{BB}, q}[m]\|_F^2 = N_s,\\
        & m = 0, 1, \cdots, M - 1.
    \end{split}
\end{equation}
where $\eta = 0$ or $\eta = 1$ stands for either sensing-only or communication-only hybrid beamforming design problem. Without loss of generality, we can consider solving the hybrid precoding problem at different time slots separately. Then, a common approach is to use alternating minimization techniques~\cite{liu2019jsc, wang2022isac, yu2016altmin}, i.e., alternately solving for $\mathbf{F}_{\text{RF}, q}$ and $\mathbf{F}_{\text{BB}, q}[m]$. Hereby, with the irregular structure of the DAoSA analog precoder, we propose an ISAC hybrid precoding algorithm by utilizing the vectorization (VEC) operation in~\cite{yan2020DAoSA}.

\subsection{VEC-based ISAC Hybrid Precoding Algorithm}

\subsubsection{Digital Precoding Design}
When fixing the analog precoder, we can impose an orthogonal constraint that $\mathbf{F}_{\text{BB}, q}[m]$ is unitary to mitigate the interference among data streams. Then, the problem \eqref{eq:weighted_sum} can be transferred to,
\begin{equation}\label{eq:opp}
    \begin{split}
        \min_{\mathbf{F}_{\text{BB}, q}[m]} & \frac{1}{M}\sum_{m=0}^{M-1} \|\mathbf{G}_q[m] - \mathbf{B}_q \mathbf{F}_{\text{BB}, q}[m]\|_F^2 \\
        \text{s.t. \quad } & \mathbf{F}_{\text{RF}, q} \in \mathcal{F},
        \mathbf{F}_{\text{BB}, q}^H[m]\mathbf{F}_{\text{BB}, q}[m] = \mathbf{I}_{N_s},\\
        & m = 0, 1, \cdots, M - 1.
    \end{split}
\end{equation}
where
\begin{equation}
    \mathbf{G}_q[m] = [\sqrt{\eta} \mathbf{F}_c^T[m], \sqrt{1 - \eta} \mathbf{F}_{s, q}^T ]^T,
\end{equation}
\begin{equation}
    \mathbf{B}_q = [\sqrt{\eta} \mathbf{F}_{\text{RF}, q}^T, \sqrt{1 - \eta} \mathbf{F}_{\text{RF}, q}^T ]^T.
\end{equation}
Similar to the solution of the so-called Orthogonal Procrustes problem (OPP)~\cite{yu2016altmin}, the solution to \eqref{eq:opp} is given by,
\begin{equation}\label{eq:digital}
    \mathbf{F}_{\text{BB}, q}[m] = \mathbf{V}_1 \mathbf{U}^H,
\end{equation}
where $\mathbf{G}_q^H[m] \mathbf{B}_q = \mathbf{U} \mathbf{\Sigma} \mathbf{V}^H$ is the SVD of $\mathbf{G}_q^H[m] \mathbf{B}_q$, and $\mathbf{V}_1$ is the first $N_s$ columns of $\mathbf{V}$.

\subsubsection{Analog Precoding Design}
When fixing the digital precoder, we carry the vectorization process and the analog precoding design problem can be formulated as,
\begin{equation}
\begin{split}
        \min_{\mathbf{F}_{\text{RF}, q}}& \frac{1}{M} \sum_{m=0}^{M-1} \bigg(\eta \|\text{vec}(\mathbf{F}_c[m]) - \text{vec}(\mathbf{F}_{\text{RF}, q}\mathbf{F}_{\text{BB}, q}[m])\|_2^2 + \\
        & (1 - \eta) \|\text{vec}(\mathbf{F}_{s, q}) - \text{vec}(\mathbf{F}_{\text{RF}, q}\mathbf{F}_{\text{BB}, q}[m])\|_2^2 \bigg).
\end{split}
\end{equation}
After removing the zero elements in $\text{vec}(\mathbf{F}_{\text{RF}, q})$, we need to solve its non-zero part $\mathbf{f}_{\text{eff}} \in \mathbb{C}^{N_{c} K_t \times 1}$, where $N_c$ denotes the number of closed switches. This is a phase rotation problem, whose solution is given by
\begin{equation}\label{eq:analog}
\begin{split}
        \mathbf{f}_\text{eff} =& \exp \Bigg(j \arg \Bigg\{\sum_{m=0}^{M-1} \mathbf{D}^H \text{vec}\biggl(\eta \mathbf{F}_c[m] \mathbf{F}_{\text{BB}, q}^H[m] \\
        &+ (1 - \eta) \mathbf{F}_{s, q} \mathbf{F}_{\text{BB}, q}^H[m]\biggr) \Bigg\} \Bigg),
\end{split}
\end{equation}
where $\mathbf{D}$ equals to $\mathbf{I}_{N_t N_\text{RF}^t}$ with $d_1$\textsuperscript{th}, $\cdots$, $d_{N_t N_\text{RF}^t - N_c K_t}$\textsuperscript{th} columns punctured, which correspond to the indices of zero elements in $\text{vec}(\mathbf{F}_{\text{RF}, q})$. Based on $\mathbf{f}_\text{eff}$, the effective analog precoder $\mathbf{F}_{\text{RF}, q}$ can be recovered. With \eqref{eq:digital} and \eqref{eq:analog}, we can alternatively calculate $\mathbf{F}_{\text{BB}, q}[m]$ and $\mathbf{F}_{\text{RF}, q}$ until convergence. After that, we finally update the digital precoders as
\begin{equation}
    \mathbf{F}_{\text{BB}, q}[m] = \frac{\sqrt{N_s}}{\|\mathbf{F}_{\text{RF}, q} \mathbf{F}_{\text{RF}, q}^{\dagger} \mathbf{G}_q[m] \|_F} \mathbf{F}_{\text{RF}, q}^{\dagger} \mathbf{G}_q[m].
\end{equation}

While the VEC algorithm provides a satisfactory solution, it requires a number of iterations in each time slot. Nevertheless, the optimal communication precoder $\mathbf{F}_c[m]$ remains the same at different time slots during a frame duration, while only the optimal sensing precoder $\mathbf{F}_{s, q}$ changes. Motivated by this, we can calculate the initial solutions of analog and digital precoders from $\mathbf{F}_c[m]$ and then update the analog precoders only once at each time slot based on the sensing codebook. Thus, we further propose the following low-complexity sensing codebook-assisted (SCA) ISAC hybrid precoding algorithm.

\subsection{Low-Complexity SCA Algorithm}
Instead of using the weighted objective function in \eqref{eq:weighted_sum}, we can define a weighted ISAC precoder as, $\mathbf{F}_{q}[m] = \beta (\sqrt{\eta} \mathbf{F}_{c}[m] + \sqrt{1 - \eta} \mathbf{F}_{s, q})$ with a normalized factor of $\beta = \sqrt{N_s} / \|\sqrt{\eta} \mathbf{F}_{c}[m] + \sqrt{1 - \eta} \mathbf{F}_{s, q}\|_F$. Before designing the ISAC analog and digital precoders, we can first obtain the solution of analog precoder for the communication-only hybrid precoding design problem,
\begin{equation}\label{eq:com}
    \begin{split}
        \overline{\mathbf{F}}_\text{RF} = \arg\min_{\mathbf{F}_{\text{RF}}, \mathbf{F}_{\text{BB}}[m]} & \frac{1}{M}\sum_{m=0}^{M-1}\|\mathbf{F}_c[m] - \mathbf{F}_{\text{RF}}\mathbf{F}_{\text{BB}}[m]\|_F^2\\
        \text{s.t. \quad } & \mathbf{F}_{\text{RF}} \in \mathcal{F},
        \|\mathbf{F}_{\text{RF}} \mathbf{F}_{\text{BB}}[m]\|_F^2 = N_s,\\
        & m = 0, 1, \cdots, M - 1,
    \end{split}
\end{equation}
which can be directly solved by the VEC algorithm.

Based on the initial analog precoder $\overline{\mathbf{F}}_\text{RF}$, we can update the analog precoder $\mathbf{F}_{\text{RF}, q}$ at the $q$\textsuperscript{th} time slot with the desired sensing beamforming vector $\mathbf{A}(:, q)$. Specifically, we calculate the error between the analog precoding vectors of the phase shifters with closed switches and corresponding columns of $\mathbf{A}(:, q)$ as,
\begin{equation}
\begin{split}
        E_{i , j} =& \|\mathbf{A}((i-1)K_t+1:iK_t, q) \\
        &- \overline{\mathbf{F}}_\text{RF}((i-1)K_t+1:iK_t, j)\|_2,
\end{split}
\end{equation}
for all $(i, j)$ satifying $\mathbf{p}_{i,j} = \mathbf{1}_{K_t}$. Then, we find the first $K_s$ minimum values of $E_{i, j}$ with the indices $\{(i_1, j_1), \cdots, (i_{K_s}, j_{K_s})\}$, where $K_s = \lceil N_c (1-\eta)\rceil$ denotes the number of subarray beamforming vectors that need to be updated. Next, we can set the designed analog precoder $\mathbf{F}_{\text{RF}, q} = \overline{\mathbf{F}}_{\text{RF}}$ and update it as,
\begin{equation}
    \mathbf{F}_{\text{RF}, q}((i_{k}-1)K_t+1:i_k K_t, j_k) = \mathbf{A}((i_k-1)K_t+1:i_k K_t, q)
\end{equation}
for $k = 1, \cdots, K_s$. The digital precoders are calculated as
\begin{equation}
    \mathbf{F}_{\text{BB}, q}[m] = \frac{\sqrt{N_s}}{\|\mathbf{F}_{\text{RF}, q} \mathbf{F}_{\text{RF}, q}^{\dagger} \mathbf{F}_q[m] \|_F} \mathbf{F}_{\text{RF}, q}^{\dagger} \mathbf{F}_q[m].
\end{equation}

\section{Sensing Estimation Algorithm Design with DAoSA Hybrid Beamforming}\label{sec:sensing_algorithm}

In this section, we propose the sensing parameter estimation algorithms at the sensing receiver. The task of the sensing receiver is to estimate the angle, range, and velocity of targets, given the transmit signal and the received sensing signal. As the whole sensing angular window is divided into $Q$ scanning directions, at the $q$\textsuperscript{th} time slot, we only sense the targets whose azimuth angles of arrival are within $-\Omega_q$, given the knowledge of the received signal $\mathbf{y}_q$ and the transmit signal $\mathbf{s}_q$.

For angle estimation, multiple signal classification (MUSIC) is a subspace-based method with super-resolution accuracy. Hereby, based on the DAoSA-MUSIC algorithm in~\cite{chen2022mimo}, we propose the wideband DAoSA-MUSIC algorithm to estimate the target angle by extending to the wideband transmission. We need to reconstruct the observation matrix by performing stacking operations on the received signals at different subcarriers. After estimating each angle parameter, we develop a range and velocity parameter estimation algorithm over two stages, i.e., sum-DFT and golden section search (S-DFT-GSS).

\subsection{W-DAoSA-MUSIC for Angle Estimation}

At the $q$\textsuperscript{th} time slot, we construct the observation vector of the sensing receiver $\mathbf{y}_q[m, n] \in \mathbb{C}^{N_\text{RF}^r \times 1}$ as,
\begin{equation}
    \begin{split}
        \mathbf{y}_q[m, n] = \mathbf{W}_{\text{RF}, q}^H \mathbf{A}_r \mathbf{S}_q[m, n] + \mathbf{E}_q[m, n],
    \end{split}
\end{equation}
where
\begin{equation}
        \mathbf{S}_q[m, n] = \mathbf{\Lambda}_q[m, n] \mathbf{A}_t^T \mathbf{F}_{\text{RF}, q} \mathbf{F}_{\text{BB}, q}[m] \mathbf{s}_q[m, n],
\end{equation}
\begin{equation}
        \mathbf{A}_r = [\mathbf{a}_r(\theta_1, \phi_1), \cdots, \mathbf{a}_r(\theta_P, \phi_P)],
\end{equation}
\begin{equation}
        \mathbf{A}_t = [\mathbf{a}_t(\theta_1, \phi_1), \cdots, \mathbf{a}_t(\theta_P, \phi_P)],
\end{equation}
\begin{equation}
    \mathbf{\Lambda}_q[m, n] = \sqrt{\frac{N_t N_r}{P}} \text{diag}\{h_1^{(q)}[m, n], \cdots, h_P^{(q)}[m, n]\},
\end{equation}
\begin{equation}
    \mathbf{E}_q[m, n] = \mathbf{W}_{\text{RF}, q}^H \mathbf{e}_q[m, n],
\end{equation}
and $h_p^{(q)}[m,n] = h_{p} e^{-j2\pi m \Delta f \tau_p} e^{j2\pi ((q - 1) T_s + n T_o) \nu_p} $. Then, we can stack all $\mathbf{y}_q[m, n]$ into one matrix as,
\begin{equation}\label{eq:Y_theta}
\begin{split}
        \mathbf{Y}_{\theta, q} =& \left[\begin{array}{ccc}
\overline{\mathbf{y}}_{q, 0}  & \ldots & \overline{\mathbf{y}}_{q, N-1}
\end{array}\right]
\end{split}
\end{equation}
with $\overline{\mathbf{y}}_{q, n} = [\mathbf{y}_q[0, n],\cdots, \mathbf{y}_q[M-1, n]]$.
The precoders and the receive steering matrix $\mathbf{A}_r$ remain the same at different symbols during a time slot. Then \eqref{eq:Y_theta} can be written as,
\begin{equation}\label{eq:music_model}
    \mathbf{Y}_{\theta, q} = \mathbf{W}_{\text{RF}, q}^H \mathbf{A}_r \overline{\mathbf{S}}_{\theta, q} + \overline{\mathbf{E}}_q,
\end{equation}
where $\overline{\mathbf{S}}_{\theta, q} = [\mathbf{S}_q[0, 0], \cdots, \mathbf{S}_q[M-1, N-1]]$ is regarded as the $P \times M N$-dimensional equivalent signal source matrix, and $\overline{\mathbf{E}}_q \in \mathbb{C}^{N_\text{RF}^r \times M N}$ refers to the noise matrix. Based on \eqref{eq:music_model}, we can perform the W-DAoSA-MUSIC algorithm to estimate the azimuth AoAs of targets.

Given the reconstructed observation matrix $\mathbf{Y}_{\theta, q}$, the covariance matrix can be calculated as,
\begin{equation}
    \mathbf{R}_{\theta, q} = \frac{1}{M N} \mathbf{Y}_{\theta, q} \mathbf{Y}_{\theta, q}^H.
\end{equation}
Then we can conduct the eigenvalue decomposition (EVD) as,
\begin{equation}\label{eq:evd}
    \mathbf{R}_{\theta, q} = \mathbf{U}_s \mathbf{\Sigma}_s \mathbf{U}_s^H + \mathbf{U}_n \mathbf{\Sigma}_n \mathbf{U}_n^H,
\end{equation}
where $\mathbf{\Sigma}_s \in \mathbb{C}^{P_q \times P_q}$ consists of $P_q$ leading eigenvalues, $\mathbf{\Sigma}_n \in \mathbb{C}^{(N_\text{RF}^r - P_q) \times (N_\text{RF}^r - P_q)}$ contains the remaining eigenvalues and $P_q$ denotes the number of targets whose azimuth AoAs are within $-\Omega_q$. With the signal subspace $\mathbf{U}_s \in \mathbb{C}^{N_\text{RF}^r \times P_q}$ and the noise subspace $\mathbf{U}_n \in \mathbb{C}^{N_\text{RF}^r \times (N_\text{RF}^r - P_q)}$, the pseudo spectrum of W-DAoSA-MUSIC can be formulated as,
\begin{equation}\label{eq:music}
    \mathbf{P}_\text{music}(\theta, \phi) = \frac{\mathbf{a}^H(\theta, \phi) \mathbf{W}_{\text{RF}, q} \mathbf{W}_{\text{RF}, q}^H \mathbf{a}(\theta, \phi)}{\mathbf{a}^H(\theta, \phi) \mathbf{W}_{\text{RF}, q} \mathbf{U}_n \mathbf{U}_n^H \mathbf{W}_{\text{RF}, q}^H \mathbf{a}(\theta, \phi)}.
\end{equation}
Finally, the AoA estimation $(\hat{\theta}_p, \hat{\phi}_p)$ can be obtained by searching the peaks of the MUSIC spectrum within the angles of $-\Omega_q$, expressed as
\begin{equation}
    (\hat{\theta}_p, \hat{\phi}_p) = \arg\max_{\theta, \phi}\mathbf{P}_\text{music}(\theta, \phi).
\end{equation}

\subsection{S-DFT-GSS for Range and Velocity Estimation}\label{sec:dft-gss}

For range and velocity estimation, the received signal model can be expressed as,
\begin{equation}
\begin{split}
        \mathbf{y}_q[m, n] =& \sum_{p=1}^P \overline{h}_p^{(q)} e^{j2\pi n T_o \nu_p} e^{-j2\pi m \Delta f \tau_p} \overline{\mathbf{x}}_{p, q}[m, n] \\
        &+ \overline{\mathbf{e}}_q[m, n],
\end{split}
\end{equation}
where
\begin{equation}
    \overline{\mathbf{x}}_{p, q}[m, n] = \mathbf{W}_{\text{RF}, q}^H \mathbf{H}_\theta(\theta_p, \phi_p) \mathbf{F}_{\text{RF}, q} \mathbf{F}_{\text{BB}, q}[m] \mathbf{s}_q[m, n],
\end{equation}
\begin{equation}
    \mathbf{H}_\theta(\theta, \phi) = \mathbf{a}_r(\theta, \phi) \mathbf{a}_t^T(\theta, \phi)
\end{equation}
\begin{equation}
    \overline{\mathbf{e}}_q[m, n] = \mathbf{W}_{\text{RF}, q}^H \mathbf{e}_q[m, n]
\end{equation}
and $\overline{h}_p^{(q)} = \sqrt{\frac{N_t N_r}{P}} h_p e^{j2\pi (q - 1) T_s \nu_p}$. For each estimated AoA parameter $(\hat{\theta}_p, \hat{\phi}_p)$, we can construct a maximum likelihood (ML) estimator by minimizing the log-likelihood function, given by
\begin{equation}
    (\hat{\tau}_p, \hat{\nu}_p) = \arg \min_{\tau, \nu, \overline{h}} \sum_{u=1}^{N_\text{RF}^r} \left\|\mathbf{Y}_{u, q} - \overline{h} \mathbf{\Psi}(\tau, \nu) \odot \hat{\mathbf{X}}_{u, q} \right\|_F^2,
\end{equation}
where
\begin{equation}
    \mathbf{Y}_{u, q} = \left[\begin{array}{ccc}
       \mathbf{y}_q(u)[0, 0]  & \ldots & \mathbf{y}_q(u)[0, N - 1]\\
       \vdots  & \ddots & \vdots\\
       \mathbf{y}_q(u)[M-1, 0]  & \ldots & \mathbf{y}_q(u)[M-1, N-1]
    \end{array}\right],
\end{equation}
\begin{equation}
    \mathbf{\Psi}(\tau, \nu) = \mathbf{\Psi}_\tau \mathbf{\Psi}_\nu^T,
\end{equation}
\begin{equation}
    \hat{\mathbf{X}}_{u, q} = \left[\begin{array}{ccc}
       \hat{\mathbf{x}}_q(u)[0, 0]  & \ldots & \hat{\mathbf{x}}_q(u)[0, N - 1]\\
       \vdots  & \ddots & \vdots\\
       \hat{\mathbf{x}}_q(u)[M-1, 0]  & \ldots & \hat{\mathbf{x}}_q(u)[M-1, N-1]
    \end{array}\right],
\end{equation}
with
\begin{equation}
    \mathbf{\Psi}_\tau = [e^{-j2\pi 0 \Delta f \tau}, e^{-j 2\pi 1 \Delta f \tau}, \cdots, e^{-j 2\pi (M - 1) \Delta f \tau}]^T,
\end{equation}
\begin{equation}
    \mathbf{\Psi}_\nu = [e^{j2\pi 0 T_o \nu}, e^{j2\pi 1 T_o \nu}, \cdots, e^{j2\pi (N - 1) T_o \nu}]^T,
\end{equation}
\begin{equation}
    \hat{\mathbf{x}}_q[m, n] =  \mathbf{W}_{\text{RF}, q}^H \mathbf{H}_\theta(\hat{\theta}_p, \hat{\phi}_p) \mathbf{F}_{\text{RF}, q} \mathbf{F}_{\text{BB}, q}[m] \mathbf{s}_q[m, n],
\end{equation}
for $u = 1, 2, \cdots, N_\text{RF}^r$. Next, this minimization problem can be transformed to the maximization problem,
\begin{equation}\label{eq:ml_estimation}
\begin{split}
        (\hat{\tau}_p, \hat{\nu}_p) = \arg \max_{\tau, \nu} \mathbf{P}_\text{ML}(\tau, \nu),
\end{split}
\end{equation}
where
\begin{equation}
\begin{split}
        \mathbf{P}_\text{ML}(\tau, \nu) &= \frac{\left|\sum_{u=1}^{N_\text{RF}^r} \text{Tr}\left(\left(\mathbf{\Psi}(\tau, \nu) \odot \hat{\mathbf{X}}_{u, q}\right)^H \mathbf{Y}_{u, q}\right)\right|^2}{\sum_{u=1}^{N_\text{RF}^r} \|\mathbf{\Psi}(\tau, \nu) \odot \hat{\mathbf{X}}_{u, q}\|_F^2} \\
        &= \left|\sum_{u=1}^{N_\text{RF}^r} \text{Tr}\left(\left(\mathbf{\Psi}(\tau, \nu) \odot \hat{\mathbf{X}}_{u, q}\right)^H \mathbf{Y}_{u, q}\right)\right|^2
\end{split}
\end{equation}
The solution in \eqref{eq:ml_estimation} is obtained by searching $(\tau, \nu)$ at which $\mathbf{P}_\text{ML}(\tau, \nu)$ achieves a maximum value in the region $[0, \frac{1}{\Delta f})\times [-\frac{1}{2T_o}, \frac{1}{2 T_o})$.

To reduce the computational complexity, we can design a two-phase estimation method. Specifically, in the first phase, we operate the on-grid search within a discretized set of delay and Doppler axes with step sizes$\frac{1}{M\Delta f}$ and $\frac{1}{N T_o}$, which can be implemented with the 2D DFT algorithm. In the second phase, based on the coarse estimation result, we conduct the off-grid estimation by introducing a 2D golden section search (GSS) method. We describe the proposed S-DFT-GSS estimation method in the following.

\subsubsection{Phase I}

To compute the ML estimator in \eqref{eq:ml_estimation}, we first perform an on-grid search on the discretized grid $\mathbf{\Gamma} = \{(\frac{m_0}{M \Delta f}, \frac{n_0}{N T_o}), m_0 = 0, \cdots, M - 1, n_0 = -\frac{N}{2}, \cdots, \frac{N}{2}-1\}$, as

\begin{equation}
    (\hat{m}_0,  \hat{n}_0) = \arg \max_{(\tau, \nu)\in \mathbf{\Gamma}} \mathbf{P}_\text{ML}\left(\frac{m_0}{M \Delta f}, \frac{n_0}{NT}\right).
\end{equation}
Hereby, we need to calculate the $M\times N$-dimensional ML estimator profiles on $\mathbf{\Gamma}$, which can be computed from the sum of $N_\text{RF}^r$ 2D DFT outputs, given by
\begin{equation}
    \mathbf{P}_\text{ML}\left(\frac{m_0}{M \Delta f}, \frac{n_0}{NT}\right) = \left|\mathbf{g}_d\left(m_0 + 1, [n_0]_N + 1\right)\right|^2
\end{equation}
where
\begin{equation}
    \mathbf{g}_d = \sum_{u=1}^{N_\text{RF}^r} \mathbf{F}_M^H \left(\hat{\mathbf{X}}_{u, q}^* \odot \mathbf{Y}_{u, q} \right) \mathbf{F}_N,
\end{equation}
and $\mathbf{F}_M\in \mathbb{C}^{M\times M}$ and $\mathbf{F}_N \in \mathbb{C}^{N \times N}$ refer to the normalized DFT matrices. Then we determine that the delay parameter lies between $\frac{\hat{m}_0 - 1}{M \Delta f}$ and $\frac{\hat{m}_0 + 1}{M \Delta f}$ and the Doppler parameter is between $\frac{\hat{n}_0 - 1}{N T_o}$ and $\frac{\hat{n}_0 + 1}{N T_o}$. Thus, the search region $\mathbf{\Gamma}_g$ for off-grid estimation in the second phase becomes,
\begin{equation}
     \left\{(\tau, \nu), \frac{\hat{m}_0 - 1}{M \Delta f}\leq \tau \leq \frac{\hat{m}_0 + 1}{M \Delta f}, \frac{\hat{n}_0 - 1}{N T_o} \leq \nu \leq \frac{\hat{n}_0 + 1}{N T_o} \right\}.
\end{equation}

\subsubsection{Phase II}
In this phase, we perform an off-grid search over the continuous-valued region $\mathbf{\Gamma}_g$, as
\begin{equation}
\begin{split}
        (\hat{\tau}_p, \hat{\nu}_p) = \arg \max_{(\tau, \nu)\in \mathbf{\Gamma}_g} \mathbf{P}_\text{ML}(\tau, \nu).
\end{split}
\end{equation}
Hereby, we can utilize the 2D golden section search technique, each step of which reduces the interval of uncertainty by
the golden ratio. Finally, the estimated velocity and range are given by $\hat{r}_p = \frac{\hat{\tau}_p c_0}{2}$ and $\hat{v}_p = \frac{\hat{\nu}_p c_0}{2 f_c}$, respectively.

\section{ISI- and ICI-tackled Sensing Algorithm}\label{sec:isi_ici}

In the previous section, the proposed estimation algorithm is based on the assumption that the round-trip delay of targets is not longer than the CP duration and the Doppler shifts are much smaller than the subcarrier spacing, i.e., the sensing channel is both \textit{ISI- and ICI-free}. Nevertheless, when it comes to the THz band, this assumption might become invalid in some cases. First, as the carrier frequency increases, the Doppler shift in the THz band grows much larger than the microwave band, which may cause inter-carrier interference and degrade sensing accuracy, especially in high-mobility scenarios. Second, with the decrease of communication delay spread in the THz band, larger subcarrier spacing can be used and the symbol and CP durations are reduced. However, this limits the maximum sensing distance if still using the proposed ISI- and ICI-unaware sensing algorithm in Sec.~\ref{sec:dft-gss} even when the link budget is sufficient.

In this section, we first derive the received signal model with ISI and ICI caused by the sensing channel and then develop an \textit{ISI- and ICI-tackled} sensing algorithm to overcome the estimation problem with ICI and ISI. Since we take into account the ISI and ICI effects, we focus on the time-frequency domain signal model and design, by simplifying the notations of the spatial domain in this section.

\subsection{Received Signal Model with ICI and ISI}

During a time slot, we denote the data signal at the $m$\textsuperscript{th} subcarrier and the $n$\textsuperscript{th} symbol as $X_{m, n}$. Then, the transmit baseband signal with the CP part is expressed as,
\begin{equation}
    s(t) = \sum_{m=1}^{M-1} \sum_{n=0}^{N-1} X_{m, n} \text{rect}\left(t - n T_o\right) e^{j 2\pi m \Delta f (t - T_\text{cp} - n T_o)},
\end{equation}
where $\text{rect}(t)$ refers to a rectangular pulse that is limited to $[0, T_o]$. At the sensing receiver, the baseband time-domain continuous signal $r(t)$ is given by,
\begin{equation}
    r(t) = \sum_{p=1}^{P} \alpha_p e^{j2\pi \nu_p t} s(t - \tau_p) + w(t),
\end{equation}
where $\alpha_p $ stands for the channel coefficient of the $p$\textsuperscript{th} target, $w(t)$ denotes the AWGN, delay and Doppler parameters are described in Sec.~\ref{sec:sensing_model} with relaxing the assumptions $\tau_p \leqslant T_\text{cp}$ into $\tau_p \leqslant T_s$ and $\nu_p \ll \Delta f $ into $\nu_p < \Delta f$. By sampling the received signal and removing the CP part, we obtain the baseband time-domain discrete signal,
\begin{equation}
    \begin{split}
        \overline{r}_{m, n} =& r(t)|_{t = nT_o + T_\text{cp} + \frac{m}{M}T} \\
        =& \sum_{p=1}^{P} \alpha_p e^{j2 \pi \nu_i (n T_o + T_\text{cp} + \frac{m}{M} T)} s\left(nT_o + T_\text{cp} + \frac{m}{M} T - \tau_p\right) \\
        &+ w_{m, n}.
    \end{split}
\end{equation}
Hereby, the key step is to derive the sampling signal $s_{\tau_p, m, n} = s\left(nT_o + T_\text{cp} + \frac{m}{M} T - \tau_p\right)$ as
\begin{equation}
    \begin{split}
        & \sum_{m'=0}^{M-1} \sum_{n'=0}^{N-1} X_{m', n'} \text{rect}\left((n - n')T_o + T_{\text{cp}} + \frac{m}{M}T - \tau_p\right) \\
        &\times e^{j 2\pi m' \Delta f \left((n - n') T_o + \frac{m}{M}T - \tau_p \right)}.
    \end{split}
\end{equation}
When $k_p T_o \leqslant \tau_p < k_p T_o + T_\text{cp} $ with $k_p = \lfloor \frac{\tau_i}{T_o} \rfloor$ ($\lfloor \cdot \rfloor$ stands for the floor function), we can obtain
\begin{equation}\label{eq:staup}
    \begin{split}
        s_{\tau_p, m, n} = \sum_{m'=0}^{M-1} X_{m', n-k_p} e^{j2\pi \frac{m' m}{M} } e^{-j2\pi m' \Delta f \tau_p} e^{j2\pi m' k_p \frac{M_\text{cp}}{M}}.
    \end{split}
\end{equation}
When $k_p T_o + T_\text{cp} \leqslant \tau_p < (k_p + 1) T_o$, for $m \geqslant \frac{\tau_p}{T}M - M_\text{cp} - k_p(M+M_\text{cp})$, $s_{\tau_p, m, n}$ is the same as that in~\eqref{eq:staup}. For $m < \frac{\tau}{T} M - M_\text{cp} - k_p (M + M_\text{cp})$, we obtain $s_{\tau_p, m, n}$ as
\begin{equation}
    \begin{split}
        \sum_{m'=0}^{M-1} X_{m', n-k-1} e^{j2\pi \frac{m' m}{M}} e^{-j2\pi m' \Delta f \tau} e^{j2\pi m' k \frac{T_\text{cp}}{T}} e^{j2\pi m' \frac{M_\text{cp}}{M}}.
    \end{split}
\end{equation}

Based on the above derivations, we can derive the time-domain input-output relation, i.e., the vector form of the received signal time-domain $\overline{r}_{m, n}$ at the $q$ time slot, $\mathbf{\overline{r}}_q \in \mathbb{C}^{MN\times 1}$, is expressed as,
\begin{equation}
    \begin{split}
        \mathbf{\overline{r}}_q =& \sum_{p=1}^{P} \alpha_p \mathbf{\Delta}^{(\nu_p)} \mathbf{D}_{N} \mathbf{\Pi}_{2MN}^{l_p + k_p M} \text{vec}\Big( \mathbf{\Pi}_M^{-l_p} (\overline{\mathbf{D}}_{l_p} \mathbf{\Pi}_M^{-M_\text{cp}} + \hat{\mathbf{D}}_{l_p}) \\
        &\cdot \mathbf{F}_M^H \mathbf{b}_{\tau_p} [\mathbf{X}_{q-1}, \mathbf{X}_q ] \Big) + \mathbf{w}_q,
    \end{split}
\end{equation}
where $\mathbf{X}_q \in \mathbb{C}^{M\times N}$ denotes the time-frequency domain transmit signal at the $q$\textsuperscript{th} time slot, $l_p = \max\{0, \lceil \frac{\tau_p}{T} M - M_\text{cp} - k_p(M + M_\text{cp}) \rceil \}$ ($\lceil \cdot \rceil$ describes the ceiling function), $\mathbf{\Delta}^{(\nu_p)} = \text{diag}(\text{vec}(\mathbf{V}_{\nu_p}))$ with $\mathbf{V}_{\nu_p}(m, n) = e^{j2\pi \nu_p (n T_o + T_\text{cp} + \frac{m}{M} T)}$, the matrix $\mathbf{\Pi}_{M} \in \mathbb{C}^{M\times M}$ refers to the forward cyclic-shift (permutation) matrix, $\mathbf{D}_{N}$ equals to the identity matrix $\mathbf{I}_{2MN}$ with the first $MN$ rows punctured, $\overline{\mathbf{D}}_{l_p}$ equals to the identity matrix $\mathbf{I}_M$ with the last $M - l_p$ rows turning into zero elements, $\hat{\mathbf{D}}_{l_p}$ equals to the identity matrix $\mathbf{I}_M$ with the first $l_p$ rows becoming zero elements, $\mathbf{b}_{\tau_p} = \text{diag}\{b_{\tau_p}^{0}, \cdots, b_{\tau_p}^{M-1}\}$ with $b_{\tau_p} = e^{j2\pi \left(k_p\frac{T_\text{cp}}{T} - \frac{\tau_p}{T}\right)}$, and $\mathbf{w}_q$ is the noise vector.

After performing DFT on the matrix form of $\mathbf{\overline{r}}_q$, $\mathbf{\overline{R}}_q = \text{vec}^{-1}(\mathbf{\overline{r}}_q) \in \mathbb{C}^{M\times N}$, we obtain the frequency-domain received signal $\mathbf{\overline{y}}_q \in \mathbb{C}^{MN\times 1}$ at the $q$\textsuperscript{th} time slot, given by
\begin{equation}\label{eq:ISI_ICI_model}
    \begin{split}
        \mathbf{\overline{y}}_q &= \text{vec}(\mathbf{F}_M \mathbf{\overline{R}}_q) \\
        &= \sum_{p=1}^{P} \alpha_p \mathbf{\overline{H}}_{p}(\tau_p, \nu_p) [\mathbf{x}_{q-1}^T, \mathbf{x}_{q}^T]^T + \mathbf{\overline{w}}_q,
    \end{split}
\end{equation}
where the matrix $\mathbf{\overline{H}}_p(\tau_p, \nu_p) \in \mathbb{C}^{MN\times 2MN}$ is given by,
\begin{equation}
    \begin{split}
        \mathbf{\overline{H}}_{p}(\tau_p, \nu_p) &= (\mathbf{I}_N \otimes \mathbf{F}_M) \mathbf{\Delta}^{(\nu_p)} \mathbf{D}_{N} \mathbf{\Pi}_{2MN}^{l_p + k_p M} \Big( \mathbf{I}_{2N} \otimes \Big( \mathbf{\Pi}_M^{-l_p} \\
        &\cdot(\overline{\mathbf{D}}_{l_p} \mathbf{\Pi}_M^{-M_\text{cp}} + \hat{\mathbf{D}}_{l_p}) \mathbf{F}_M^H \mathbf{b}_{\tau_p} \Big) \Big),
    \end{split}
\end{equation}
and $\mathbf{x}_{q-1} = \text{vec}(\mathbf{X}_{q - 1})$, $\mathbf{x}_{q} = \text{vec}(\mathbf{X}_{q})$. If the ISI and ICI effects are ignored, the input-output relation in the time-frequency domain is approximated as the following matrix form,
\begin{equation}
    \mathbf{\overline{Y}}_q \approx \sum_{p=1}^P \alpha_p \mathbf{X}_q \odot \mathbf{\Psi}(\tau_p, \nu_p) + \mathbf{\overline{W}}_q.
\end{equation}
The ISI- and ICI-unaware estimation is based on this approximated input-output relation, which is not accurate and causes estimation error in the presence of ISI and ICI effects.

\subsection{ISI- and ICI-tackled Estimator}

Based on the received sensing signal model with ISI and ICI in~\eqref{eq:ISI_ICI_model}, we can obtain the ISI- and ICI-tackled estimator, given by
\begin{equation}
            (\hat{\tau}, \hat{\nu}) = \arg \max_{\tau, \nu} \frac{\left(\mathbf{\overline{H}}_{p}(\tau, \nu) [\mathbf{x}_{q-1}^T, \mathbf{x}_{q}^T]^T \right)^H \mathbf{y}_q}{\|\mathbf{\overline{H}}_{p}(\tau, \nu) [\mathbf{x}_{q-1}^T, \mathbf{x}_{q}^T]^T \|_2^2}.
\end{equation}
The complexity of the proposed ISI- and ICI-tackled estimation algorithm depends on the computation of $\mathbf{\overline{H}}_{p}(\tau, \nu) [\mathbf{x}_{q-1}^T, \mathbf{x}_{q}^T]^T$. This can be implemented with computationally efficient operations, including FFT algorithms, cyclic shift, vectorization, and Hadamard product. Thus, the overall computational complexity of this estimator is $\mathcal{O}(MN \log (MN))$.

\section{Numerical Results}\label{sec:simulation}

\begin{table}[]
    \centering
    \caption{Simulation Parameters}
    \label{tab:parameter}
    \begin{tabular}{ccc}
    \toprule
    \textbf{Notations} & \textbf{Definition} & \textbf{Value} \\
    \midrule
    $f_c$ & Carrier frequency & 0.3 THz \\
    $\Delta f$ & Subcarrier spacing & 120 to 3840 kHz \\
    $T$ & Symbol duration & 8.33 to 0.26 $\mu s$\\
    $T_\text{cp}$ & CP duration & $\frac{1}{4}T$\\
    $M$ & Number of subcarriers & \{64, 1024\}\\
    $N$ & Number of symbols during a time slot & 16\\
    $Q$ & Number of time slots during a frame & 32 \\
    $N_t$ & Number of transmit antennas & 1024 \\
    $N_r$ & Number of receive antennas & 1024 \\
    $W$ & Number of antennas on the $y$-axis & 32 \\
    $L$ & Number of antennas on the $z$-axis & 32 \\
    $N_\text{RF}^t$ & Number of transmit RF chains & 4\\
    $N_\text{RF}^r$ & Number of receive RF chains & 4\\
    $N_s$ & Number of data streams & 4\\
    $N_c$ & Number of closed switches & \{4, 8, 16\} \\
    \bottomrule
    \end{tabular}
\end{table}

In this section, we evaluate the sensing and communication performance of the proposed precoding algorithms and sensing parameter estimation methods. The key simulation parameters are listed in Table~\ref{tab:parameter}, which refer to the physical layer numerology for beyond 52.6 GHz communications in~\cite{levanen2021waveform} and the THz link budget analysis in~\cite{rikkinen2020THz, Akyildiz2022THz}. We consider a THz multipath channel with one LoS path and $L_\text{N}$ = 4 NLoS paths.
In the simulations, we consider 2D beamforming, i.e., all elevation angles are set as $\phi_0 = 90^{\circ}$.

\subsection{Performance of Hybrid Precoding Algorithms for THz ISAC}

\begin{figure}
    \centering
    \includegraphics[width=0.43\textwidth]{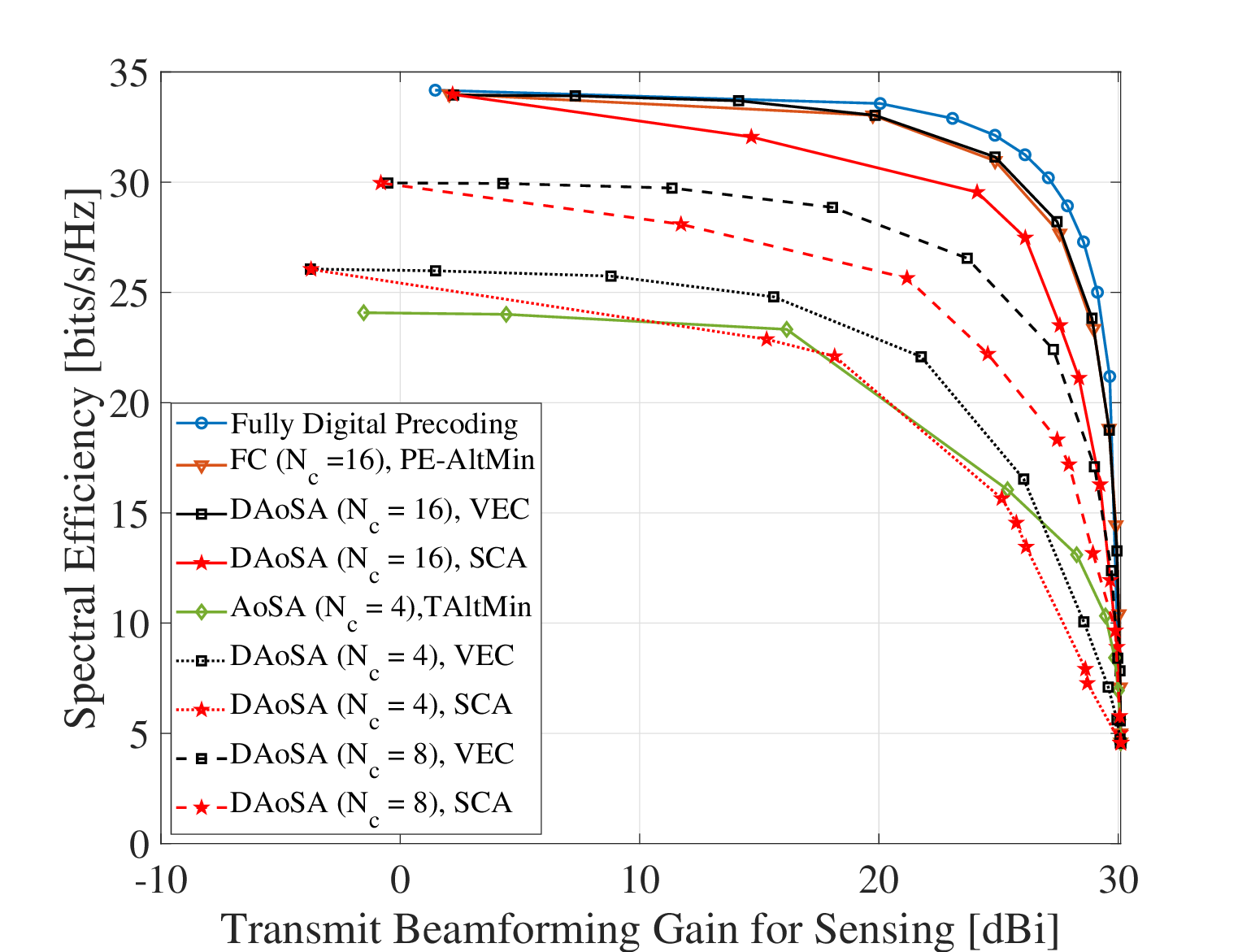}
    \caption{Performance tradeoff between spectral efficiency and transmit sensing beamforming gain when the signal-to-noise ratio (SNR) $\frac{\rho}{\sigma_n^2}$ = -20 dB.}
    \label{fig:ISAC_tradeoff}
\end{figure}

\begin{figure}
    \centering
    \includegraphics[width=0.43\textwidth]{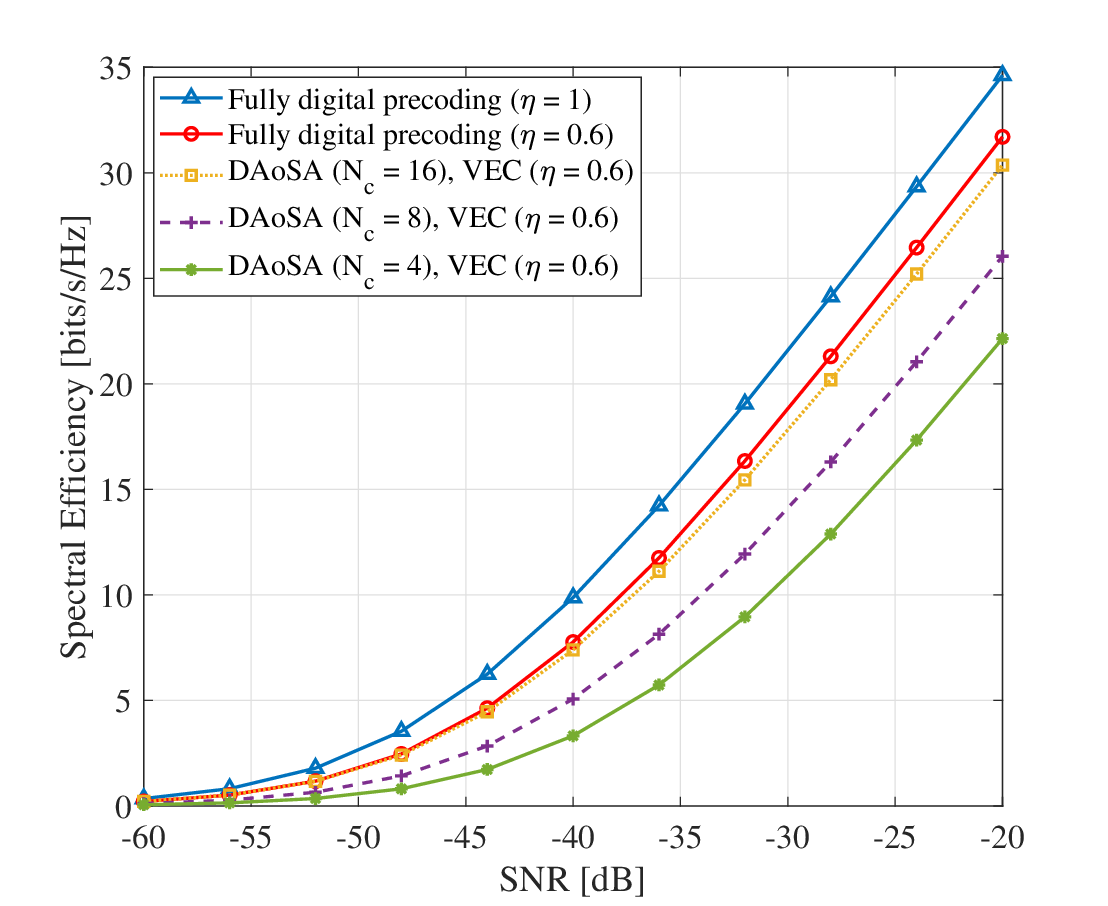}
    \caption{Spectral efficiency versus SNR for the communication link.}
    \label{fig:SE_versus_SNR}
\end{figure}

First, we evaluate the performance of the proposed VEC and SCA hybrid precoding algorithms for THz ISAC in terms of spectral efficiency and transmit beamforming gain towards the sensing direction. Specifically, we consider three hybrid precoding architectures, i.e., FC, AoSA, and DAoSA structures. In comparison, the PE-AltMin approach~\cite{yu2016altmin} and the TAltMin~\cite{liu2019jsc} algorithm are used for the FC and the AoSA structures, respectively. The proposed VEC and SCA algorithms are performed for the DAoSA architecture, which is equivalent to FC with $N_c = {(N_\text{RF}^t)}^2$ and AoSA with $N_c = N_\text{RF}^t$. Since we focus on the evaluations of the hybrid precoding design, the FC combining architecture is set at the communication receiver side. Moreover, the performance of fully digital precoding is evaluated as an upper bound. The subcarrier spacing is set as 1.92 MHz and the number of subcarriers equals 64. The signal-to-noise ratio (SNR) of the communication link is -20 dB.

\begin{figure}
    \centering
    \includegraphics[width=0.43\textwidth]{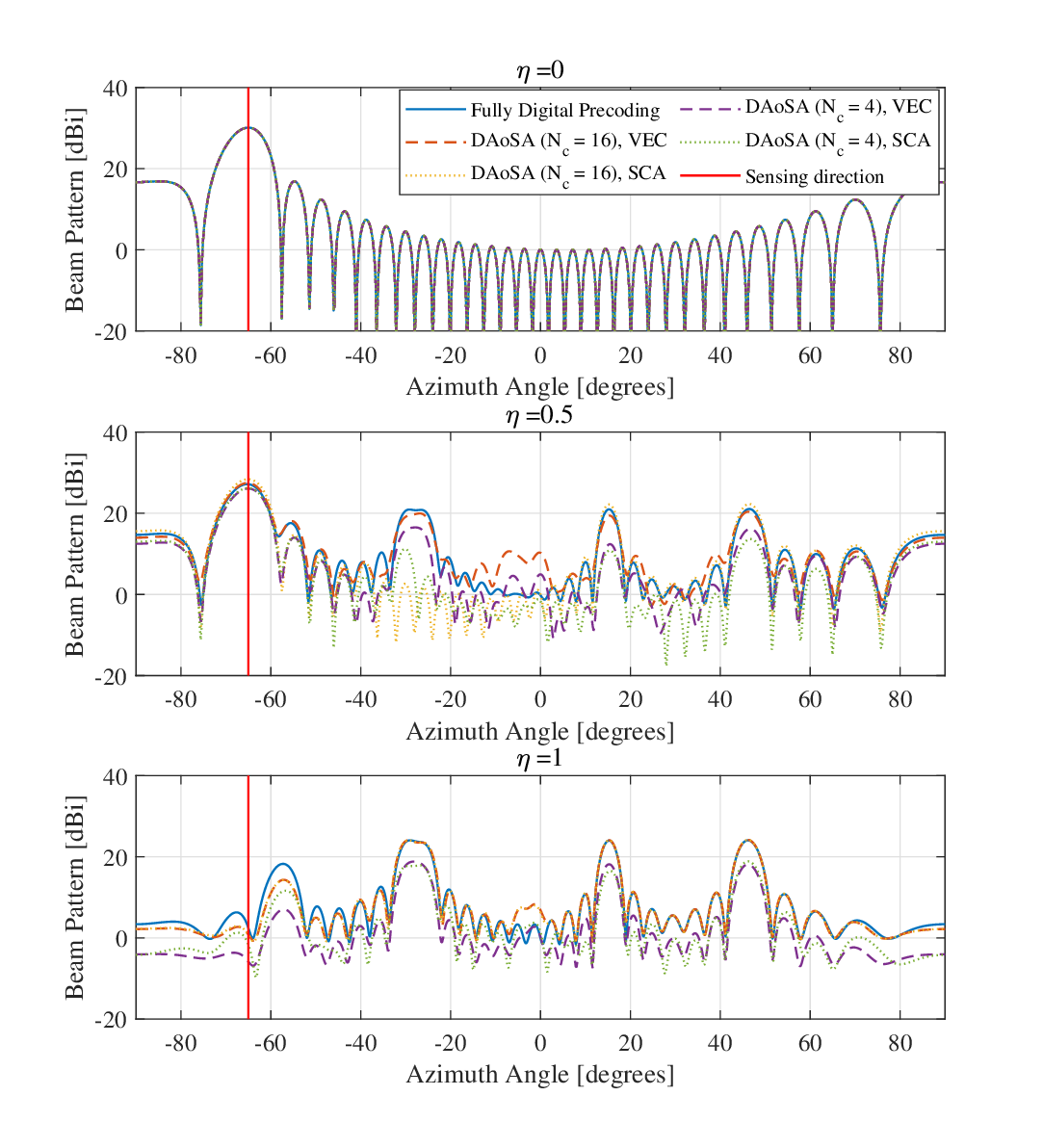}
    \caption{Transmit beampattern in the azimuth plane for $\eta = 0$, $\eta = 0.5$ and $\eta = 1$, when the sensing direction is ($-65^{\circ}, 90^{\circ}$) and the azimuth angles of communication paths are distributed between $-60^{\circ}$ and $60^{\circ}$.}
    \label{fig:ISAC_beampattern}
\end{figure}

As shown in Fig.~\ref{fig:ISAC_tradeoff}, the performance tradeoff between spectral efficiency and transmit sensing beamforming gain using different hybrid precoding algorithms is plotted by setting the weighting factor within [0, 1]. We learn that the spectral efficiency decreases as the transmit sensing beamforming gain is improved as expected, since more energy is concentrated toward the sensing direction. In the FC structure, the proposed VEC algorithm performs slightly better than the PE-AltMin approach and achieves close performance to the fully digital precoding. In the AoSA architecture, the VEC algorithm realizes higher spectral efficiency than the TAltMin method when $\eta > 0.5$, i.e., communication dominates the precoding design. Moreover, while the proposed VEC algorithm outperforms the SCA method for all dynamic hybrid beamforming structures, the SCA algorithm is more computationally efficient.

Next, we investigate the spectral efficiency versus SNR with different numbers of closed switches. In Fig.~\ref{fig:SE_versus_SNR}, compared to the communication-only precoding design ($\eta = 1$), the spectral efficiency of the ISAC precoding design ($\eta = 0.6$) is reduced by approximately 2.5 bits/s/Hz at the SNR of -30 dB. When $N_c = 16$, the DAoSA structure becomes FC, and the proposed VEC ISAC hybrid precoding algorithm achieves near-optimal performance over the whole SNR range. With fewer closed switches, fewer phase shifters are used, which causes some performance loss while improving energy efficiency.

\subsection{Transmit Beampattern}

\begin{figure}
    \centering
    \includegraphics[width=0.43\textwidth]{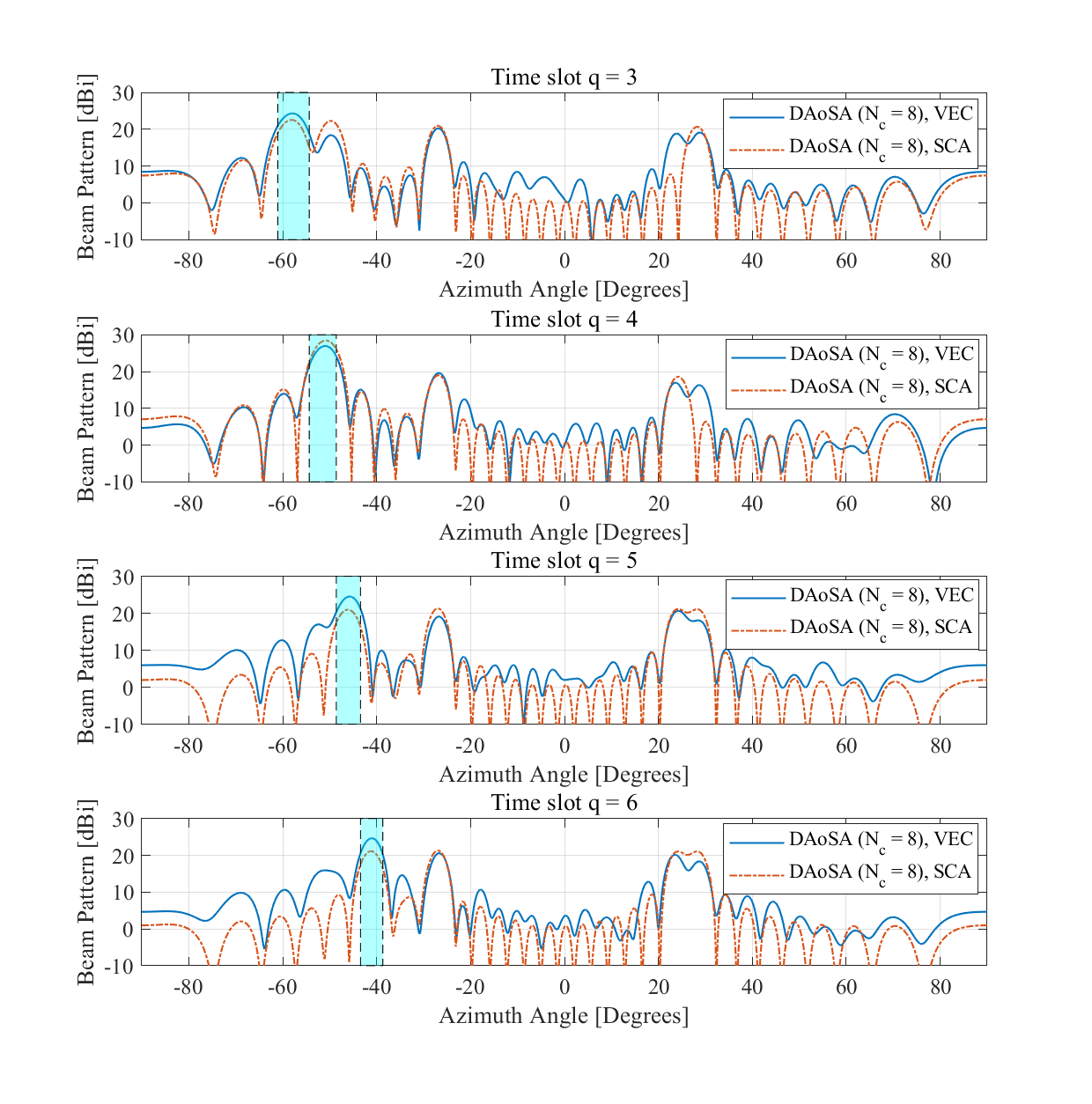}
    \caption{Beam scanning in the azimuth plane, when the transmit beampattern consists of time-varying sensing beams and stable communications beams. The sensing angular windows are depicted by the shadow areas, which can cover all angles from $-90^{\circ}$ to $90^{\circ}$.  In this plot, beam patterns at time slots $q = 3, 4, 5, 6$ are exhibited.}
    \label{fig:ISAC_beam_scanning}
\end{figure}

We illustrate the transmit beampattern of the designed hybrid precoders in Fig.~\ref{fig:ISAC_beampattern} and Fig.~\ref{fig:ISAC_beam_scanning} for different weights of ISAC precoding design and beam scanning over sequential time slots.
As shown in Fig.~\ref{fig:ISAC_beampattern}, $\eta = 0$ corresponds to the sensing-only precoder $\mathbf{F}_{s, q}$. In this case, both the proposed VEC and SCA can realize the desired beampattern in the FC ($N_c$ = 16) and AoSA ($N_c = 4$) architectures, which is generated from the DFT sensing codebook. When $\eta$ becomes 0.5, we learn that the beamforming gain toward the sensing direction is slightly reduced while several communication sub-beams are formed and point to the angles of communication paths. In the case of $\eta = 1$, the communication-only precoding design does not generate sensing beams toward the sensing direction and concentrates all beams toward the communication receiver. In addition, it is demonstrated that the transmit beam in the FC structure realizes more similar pattern to the fully digital precoding compared with the AoSA structure.

In Fig.~\ref{fig:ISAC_beam_scanning}, it is shown that during a frame duration, the designed THz ISAC transmit signal can generate sweeping beams to scan possible targets in the surrounding environment over different time slots and stable beams toward the communication user to enable ultra-fast data transmission. We observe that the transmit beamforming gains toward the sensing direction can achieve approximately 20 dBi as the beam angle varies, while the communication beams remain similar at different time slots.

\textbf{Complexity Analysis:} We denote $N_\text{iter}$ as the number of iterations of the alternating minimization in the VEC algorithm for each time slot. The overall computational complexity of the VEC-based ISAC hybrid precoding algorithm is given by $\mathcal{O}(Q N_\text{iter} N_t^2 )$.
Since the SCA ISAC hybrid precoding algorithm does not require the process of alternating minimization for each time slot, it can reduce the computational complexity to $\mathcal{O}(N_\text{iter}N_t^2)$ compared with the VEC algorithm.

\subsection{Sensing Accuracy}

\begin{figure}
    \centering
    \subfigure[]{\includegraphics[width=0.2\textwidth]{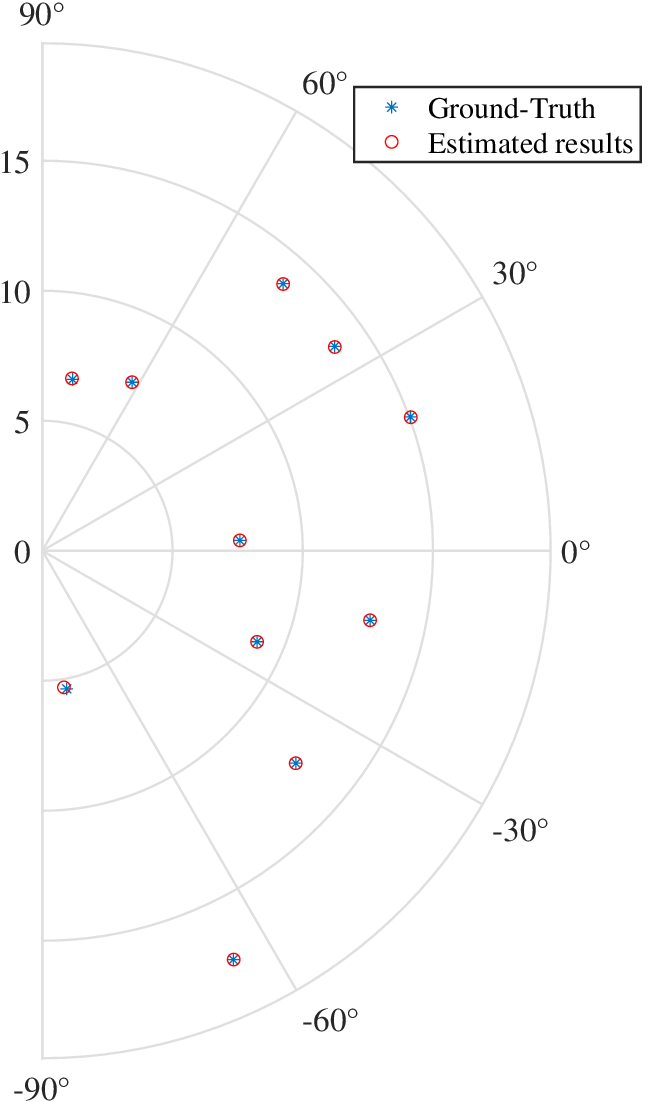}}
    \subfigure[]{\includegraphics[width=0.21\textwidth]{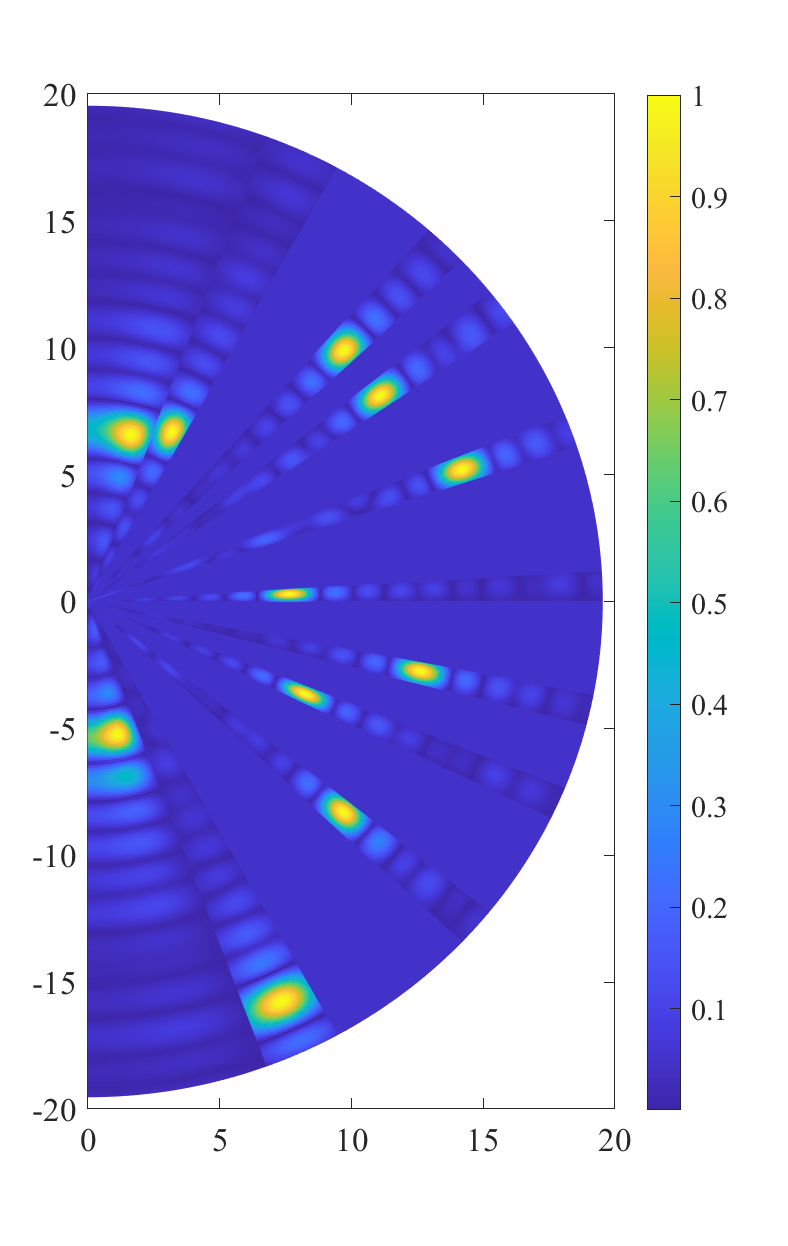}}
    \caption{Angle (degree) and distance (meter) estimation results using the proposed sensing algorithm. (a) Direct plot of the real and estimated values of the target parameters. (b) Normalized angle and range profiles of the estimates.}
    \label{fig:multi_target_estimation}
\end{figure}

\begin{figure}
    \centering
    \includegraphics[width=0.47\textwidth]{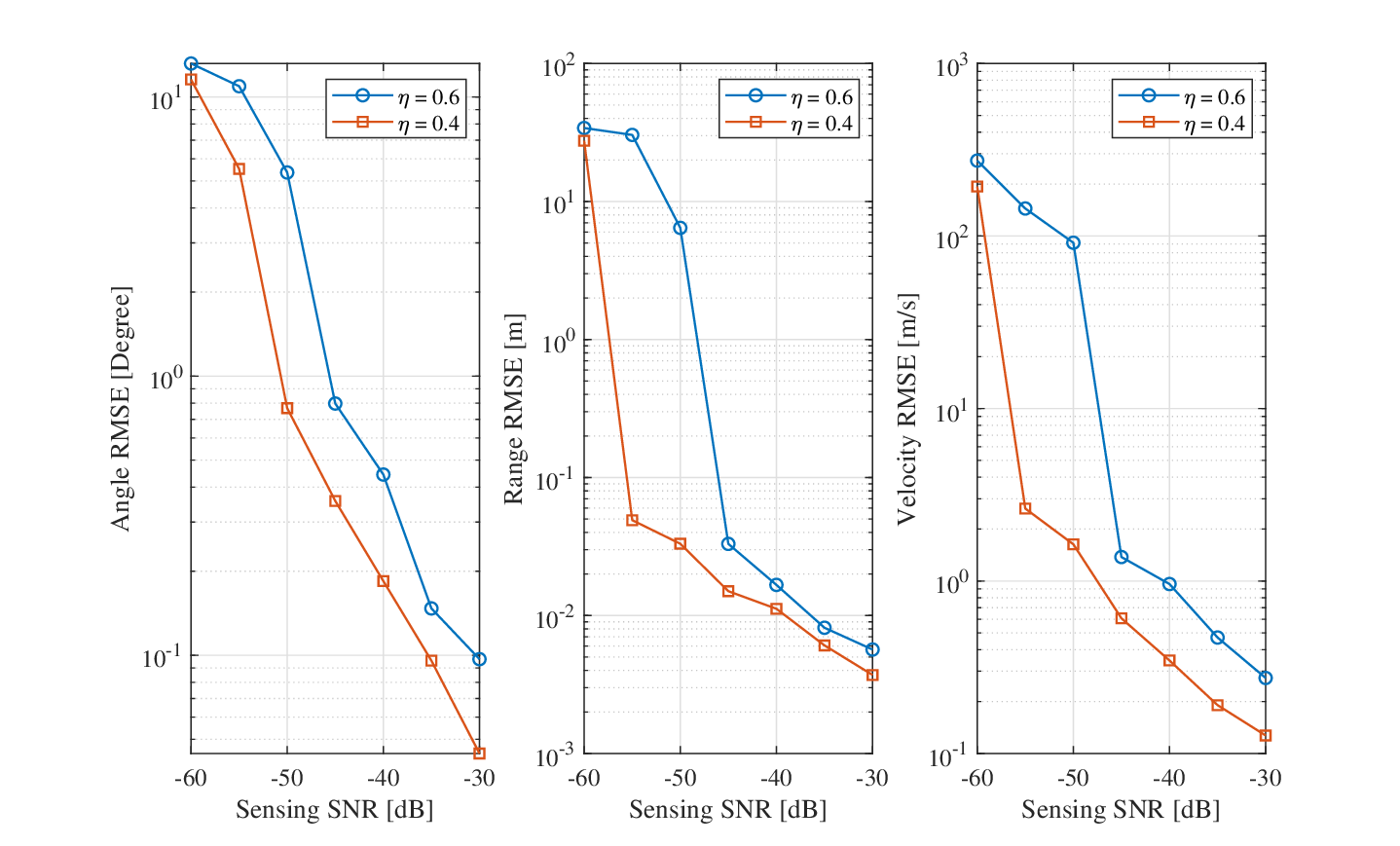}
    \caption{Root mean square error (RMSE) for angle, range, and velocity estimation on a target with the azimuth angle of $70^{\circ}$, the distance of 15 m, and the velocity of 20 m/s versus the sensing SNR.}
    \label{fig:sensing_accuracy}
    \vspace*{-10pt}
\end{figure}

We further investigate the effectiveness of the proposed sensing algorithm with the DAoSA hybrid beamforming architecture. In Fig.~\ref{fig:multi_target_estimation}, a number of sensing targets are randomly distributed between $-90^{\circ}$ and $90^{\circ}$. We conduct beam scanning by using the proposed hybrid precoding algorithms in Sec.~\ref{sec:hybrid_precoding} and then plot the normalized range profile based on the back-reflected sensing received signal by using the proposed sensing estimation algorithms in Sec.~\ref{sec:sensing_algorithm}. At the $q$\textsuperscript{th} time slot, we estimate the parameters of the target within the sensing angular window $\Omega_q$. With the time-frequency-space transmit design, we realize entire-space multi-target sensing, although the directional narrow beams are used in the THz band.

\begin{figure}
    \centering
    \subfigure[Three targets have the same velocity $v$ = 5 m/s.]{\includegraphics[width=0.43\textwidth]{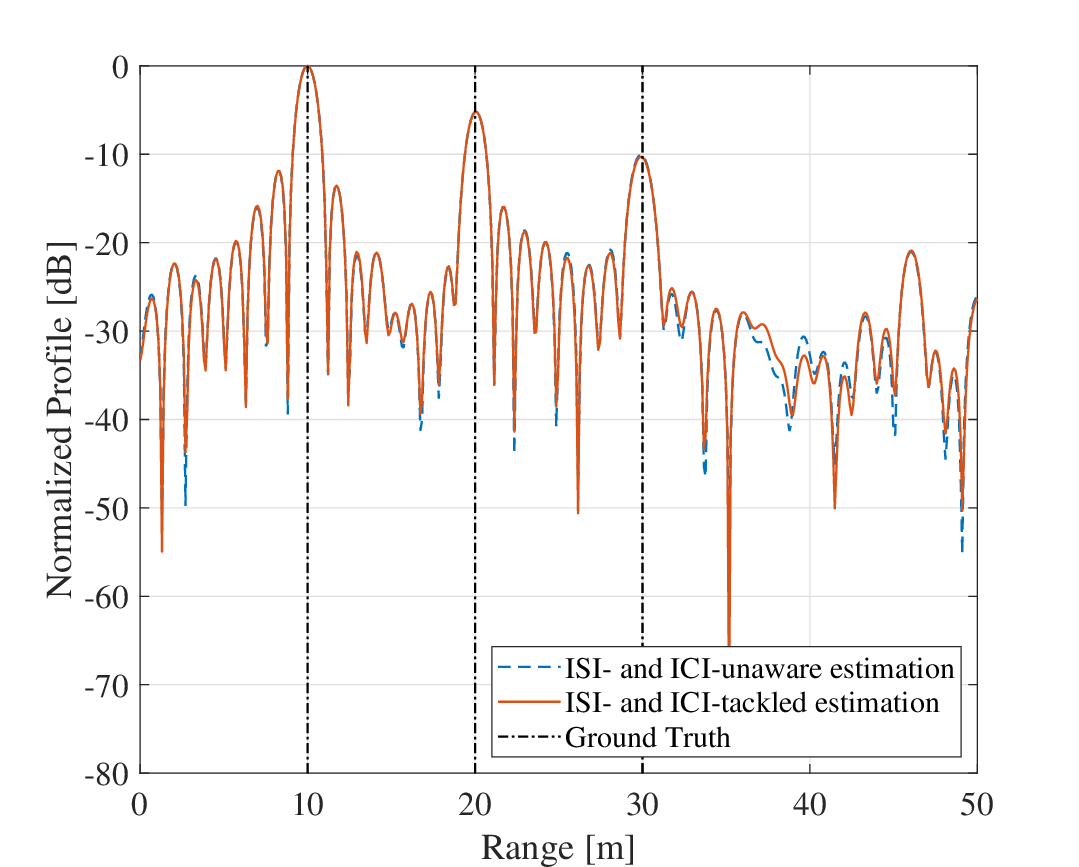}}
    \subfigure[Three targets have the same velocity $v$ = 50 m/s.]{\includegraphics[width=0.43\textwidth]{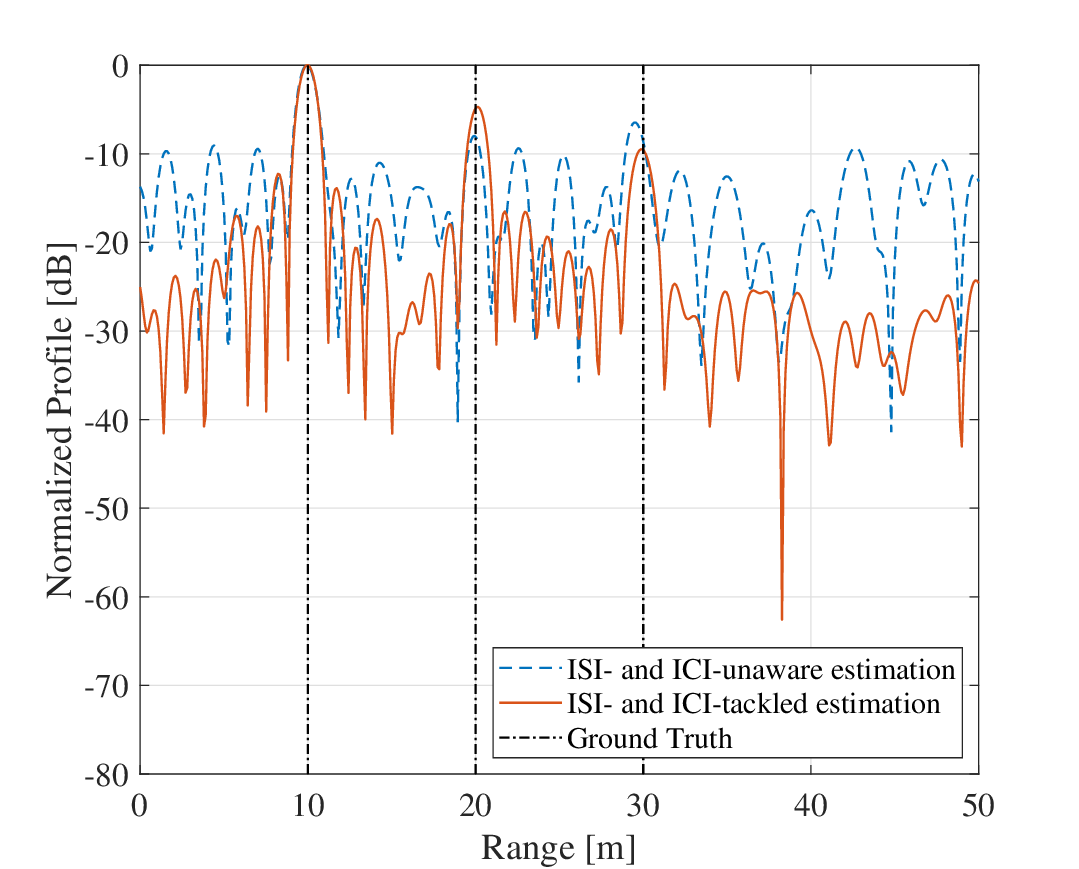}}
    \caption{Normalized range profiles of ICI-unaware and -tackled estimations through the sensing channel with weak and strong ICI effects. (a) The scenario contains 3 targets with the ranges (10, 20, 30) m and the effective SNRs (-10, -15, 20) dB considering the beamforming gain. The waveform parameters are $M$ = 1024 and $\Delta f$ = 120 kHz.}
    \label{fig:rangeRMSE_ICI}
    \vspace*{-10pt}
\end{figure}

Moreover, we evaluate the sensing accuracy of angle, range, and velocity estimation with the proposed sensing algorithm. In Fig.~\ref{fig:sensing_accuracy}, we set the target parameters including the azimuth angle of $70^{\circ}$, the distance of 15 m, and the velocity of 20 m/s. The waveform parameters are $M$ = 64 and $\Delta f$ = 3.84 MHz. The number of closed switches is 4 at both transmitter and sensing receiver sides. As the sensing SNR increases, the sensing accuracy is improved. Specifically, we observe that the angle, range, and velocity estimation can achieve centi-degree-level, millimeter-level, and decimeter-per-second-level accuracy, respectively. In addition, by decreasing the weighting factor $\eta$ from 0.6 to 0.4, the sensing accuracy is improved, since more power is allocated to the sensing beam.

\textbf{Complexity Analysis:} The computational complexity of EVD in~\eqref{eq:evd} is $\mathcal{O}((N_\text{RF}^r)^3)$. Since $N_\text{RF}^r$ is much smaller than $N_r$, the overall computational complexity of W-DAoSA-MUSIC mainly depends on the matrix-vector multiplication in~\eqref{eq:music}, namely, $\mathcal{O}(N_\text{RF}^r N_r)$. The computational complexity of the S-DFT-GSS algorithm is $\mathcal{O}(N_\text{RF}^r M N \log (MN))$ in the first phase and $\mathcal{O}(N_\text{gss} N_\text{RF}^r M N)$ in the second phase, where $N_\text{gss}$ denotes the iterations of golden section search.

\subsection{ISI and ICI Effects on Sensing Parameter Estimation}

\begin{figure}
    \centering
    \subfigure[$\Delta f$ = 480 kHz, the CP-limited maximum sensing distance is 78 m, and the sensing channel is ISI-free.]{\includegraphics[width=0.43\textwidth]{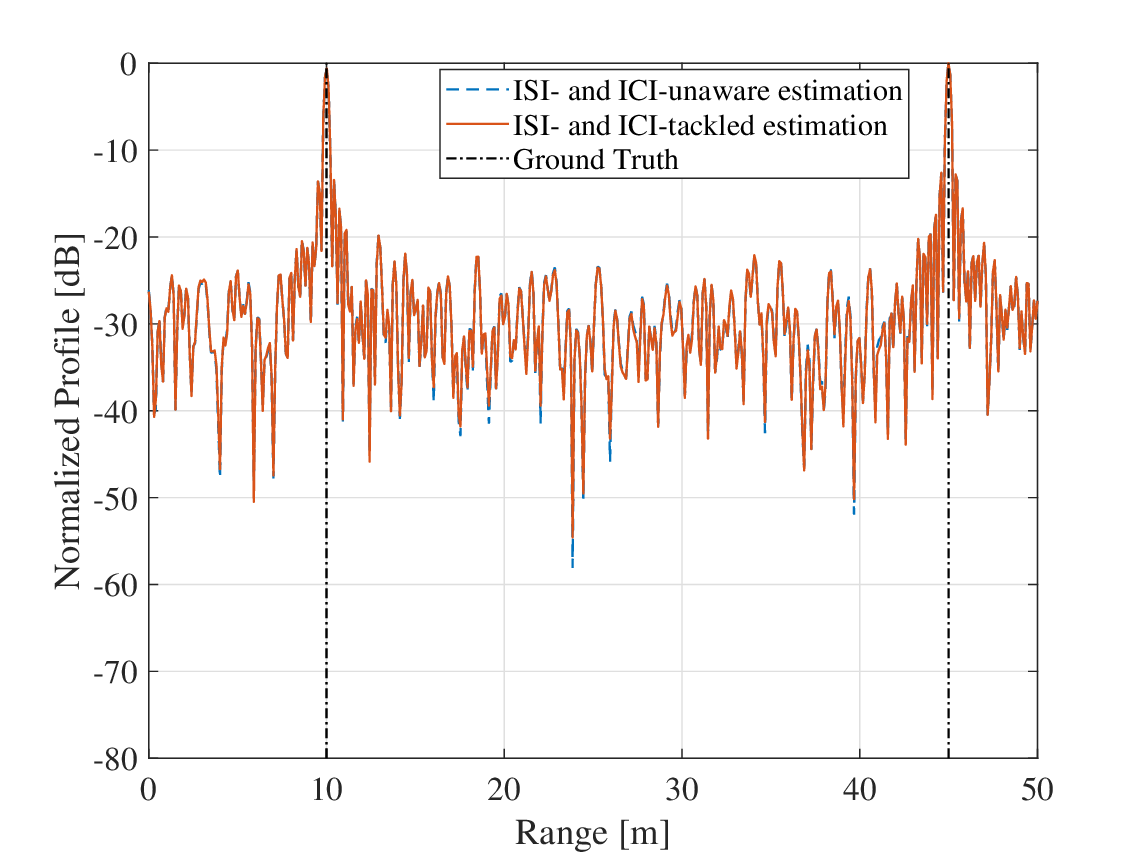}}
    \subfigure[$\Delta f$ = 3840 kHz, the CP-limited maximum sensing distance is 9.8 m, and there exist ISI effects.]{\includegraphics[width=0.43\textwidth]{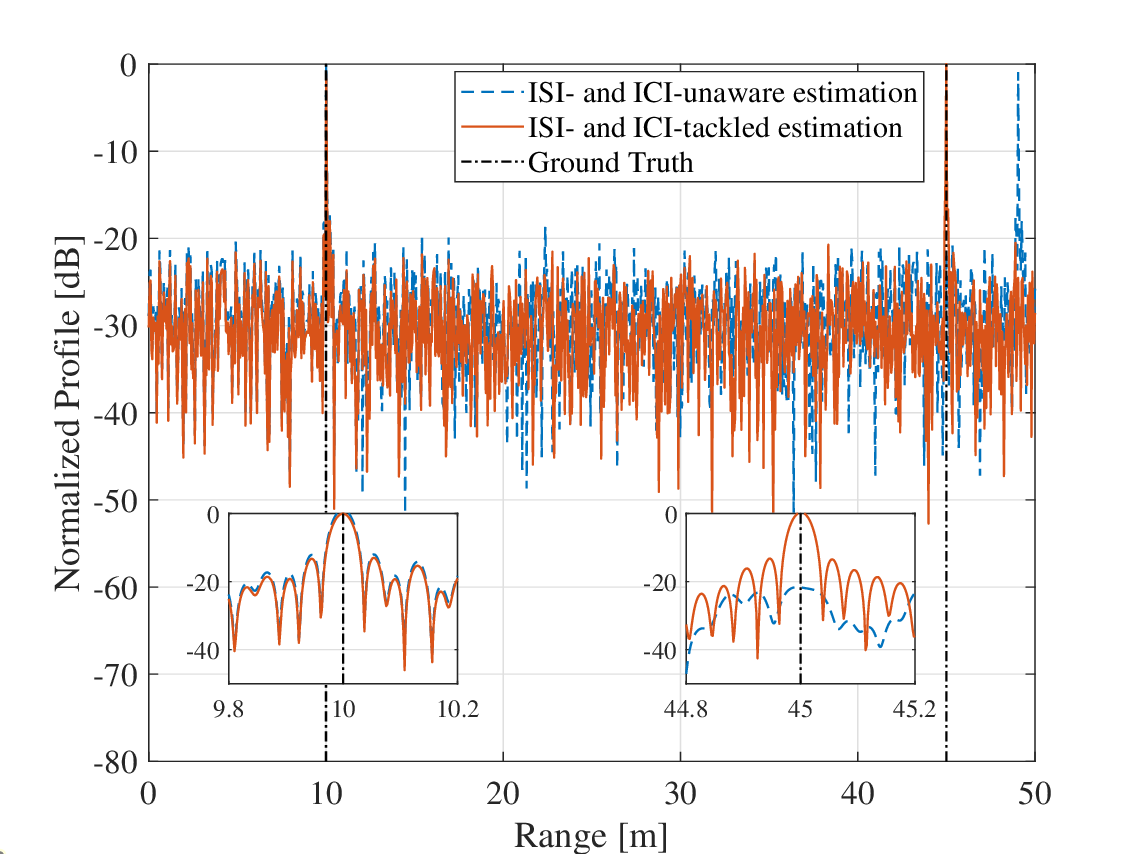}}
    \caption{Normalized range profiles of ISI-unaware and -tackled estimations through the sensing channel without and with ISI effects. (a) The scenario contains 2 targets with the ranges (10, 45) m, the same velocity $v$ = 5 m/s, and the effective SNRs (-10, -10) dB considering the beamforming gain. The subcarrier number is $M$ = 1024.}
    \label{fig:rangeRMSE_ISI}
    \vspace*{-10pt}
\end{figure}

Finally, we study the ISI and ICI effects on sensing parameter estimation for THz ISAC systems. The subcarrier number is set as 1024. The considered scenario contains 3 targets with the ranges (10, 20, 30) m and the effective SNRs (-10, -15, 20) dB considering the beamforming gain. In Fig.~\ref{fig:rangeRMSE_ICI}, we compare the ICI-unaware and ICI-tackled estimation algorithms under two cases, i.e., sensing channels with weak and strong ICI effects, respectively. As shown in Fig.~\ref{fig:rangeRMSE_ICI}(a), the velocity of targets is set as 5 m/s, which corresponds to the low-mobility scenario. In this case, we learn that both ICI-unaware and ICI-tackled sensing algorithms have similar estimation results and can accurately estimate the parameters of 3 targets. Nevertheless, when the target velocity increases to 50 m/s in Fig.~\ref{fig:rangeRMSE_ICI}(b), with ICI-unaware estimation, ICI effects increase side-lobe levels of the target with the strongest power, which may cause masking of weak targets or large errors on the parameters of the other two targets. The distance of the target at 30 m is estimated as 29.4 m and the target at 20 m cannot be detected successfully due to the ambiguity caused by side lobes. In contrast, the proposed ICI-tackled sensing algorithm can overcome this problem and still accurately estimate these three targets.

In Fig.~\ref{fig:rangeRMSE_ISI}, we consider the ISI effects on THz ISAC systems. We consider the scenario containing 2 targets with the ranges (10, 45) m, the same velocity $v$ = 5 m/s, and the effective SNRs (-10, -10) dB considering the beamforming gain. As shown in Fig.~\ref{fig:rangeRMSE_ISI}(a), when the subcarrier spacing is 480 kHz, the CP-limited maximum sensing distance is 78 m, which is longer than the target ranges. In this case, there is no ISI effect and we can obtain accurate estimated values of target ranges by using the ISI-unaware sensing algorithm. When the delay spread of the THz communication channel decreases, we can increase the subcarrier spacing and the CP duration becomes shorter, which reduces the CP-limited sensing distance. In Fig.~\ref{fig:rangeRMSE_ISI}(b), the subcarrier spacing increases to 3.84 MHz, and the CP-limited sensing distance is 9.8 m, which is shorter than the target ranges. Thus, there exist ISI effects on the received sensing signal. According to the normalized range profile using the ISI-unaware sensing algorithm, the range of the second target is estimated as 49 m, while the ground truth is 45 m. By comparison, the ISI-tackled sensing algorithm still performs well and is robust against the ISI effect.

\section{Conclusion}\label{sec:conclusion}

In this paper, we have proposed a THz ISAC system framework, including the time-frequency-space transmit design with the DAoSA hybrid beamforming architecture and OFDM waveform, and sensing algorithms for angle, range, and velocity estimation. We propose two ISAC hybrid precoding algorithms, i.e., the near-optimal VEC method and the low-complexity SCA approach. Meanwhile, in the ISI- and ICI-free case, we propose the W-DAoSA-MUSIC angle estimation algorithm and the S-DFT-GSS range and velocity estimation method. Furthermore, when there exist ISI and ICI effects on target estimation in the THz band, we develop the ISI- and ICI-tackled sensing algorithm to overcome the CP limitation and high-mobility target estimation problem.

With extensive simulations, the results indicate that the proposed VEC ISAC hybrid precoding algorithm can achieve close performance to fully digital precoding and outperforms other existing methods. The developed SCA algorithm can reduce computational complexity by removing the process of alternating minimization for each time slot. Meanwhile, with the proposed estimation algorithms, centi-degree-level angle estimation, millimeter-level range estimation, and decimeter-per-second-level velocity estimation can be realized in THz ISAC systems. Moreover, an implementation demo of OFDM waveform based THz communication system can be found in our video \textcolor{red}{\url{https://twclabsjtu.github.io/THz-ISAC-System/}}. 

\bibliographystyle{IEEEtran}
\bibliography{journal_short}

\begin{thebibliography}{10}
\providecommand{\url}[1]{#1}
\csname url@samestyle\endcsname
\providecommand{\newblock}{\relax}
\providecommand{\bibinfo}[2]{#2}
\providecommand{\BIBentrySTDinterwordspacing}{\spaceskip=0pt\relax}
\providecommand{\BIBentryALTinterwordstretchfactor}{4}
\providecommand{\BIBentryALTinterwordspacing}{\spaceskip=\fontdimen2\font plus
\BIBentryALTinterwordstretchfactor\fontdimen3\font minus \fontdimen4\font\relax}
\providecommand{\BIBforeignlanguage}[2]{{%
\expandafter\ifx\csname l@#1\endcsname\relax
\typeout{** WARNING: IEEEtran.bst: No hyphenation pattern has been}%
\typeout{** loaded for the language `#1'. Using the pattern for}%
\typeout{** the default language instead.}%
\else
\language=\csname l@#1\endcsname
\fi
#2}}
\providecommand{\BIBdecl}{\relax}
\BIBdecl

\bibitem{wu2023signal}
Y.~Wu and C.~Han, ``Time-frequency-space signal design with dynamic subarray for terahertz integrated sensing and communication,'' in \emph{Proc. of IEEE International Workshop on Signal Processing Advances in Wireless Communications (SPAWC)}, 2023.

\bibitem{Tong20216G}
W.~Tong and P.~Zhu, ``6{G}: The {N}ext {H}orizon: From connected people and things to connected intelligence,'' \emph{Cambridge University Press}, 2021.

\bibitem{Akyildiz2022THz}
I.~F. Akyildiz \emph{et~al.}, ``Terahertz band communication: An old problem revisited and research directions for the next decade,'' \emph{IEEE Transactions on Communications}, vol.~70, no.~6, pp. 4250--4285, 2022.

\bibitem{liu2022ISAC}
F.~Liu \emph{et~al.}, ``Integrated sensing and communications: Toward dual-functional wireless networks for 6{G} and beyond,'' \emph{IEEE Journal on Selected Areas in Communications}, vol.~40, no.~6, pp. 1728--1767, 2022.

\bibitem{han2023thz-isac}
C.~\vspace{0mm}Han \emph{et~al.}, ``{THz} {ISAC}: A physical-layer perspective of terahertz integrated sensing and communication,'' \emph{IEEE Communications Magazine}, vol.~62, no.~2, pp. 102--108, 2024.

\bibitem{han2021hybrid}
C.~Han \emph{et~al.}, ``Hybrid beamforming for terahertz wireless communications: Challenges, architectures, and open problems,'' \emph{IEEE Wireless Communications}, vol.~28, no.~4, pp. 198--204, 2021.

\bibitem{zhang2022jcrs}
J.~A. Zhang \emph{et~al.}, ``Enabling joint communication and radar sensing in mobile networks—a survey,'' \emph{IEEE Communications Surveys \& Tutorials}, vol.~24, no.~1, pp. 306--345, 2022.

\bibitem{han2022channel}
C.~Han \emph{et~al.}, ``Terahertz wireless channels: A holistic survey on measurement, modeling, and analysis,'' \emph{IEEE Communications Surveys \& Tutorials}, vol.~24, no.~3, pp. 1670--1707, 2022.

\bibitem{sturm2011waveform}
C.~Sturm and W.~Wiesbeck, ``Waveform design and signal processing aspects for fusion of wireless communications and radar sensing,'' \emph{Proceedings of the IEEE}, vol.~99, no.~7, pp. 1236--1259, 2011.

\bibitem{sarieddeen2021overview}
H.~Sarieddeen \emph{et~al.}, ``An overview of signal processing techniques for terahertz communications,'' \emph{Proceedings of the IEEE}, vol. 109, no.~10, pp. 1628--1665, 2021.

\bibitem{berger2010ofdm}
C.~R. Berger \emph{et~al.}, ``Signal processing for passive radar using {OFDM} waveforms,'' \emph{IEEE Journal of Selected Topics in Signal Processing}, vol.~4, no.~1, pp. 226--238, 2010.

\bibitem{johnston2022ofdm}
J.~Johnston \emph{et~al.}, ``{MIMO OFDM} dual-function radar-communication under error rate and beampattern constraints,'' \emph{IEEE Journal on Selected Areas in Communications}, vol.~40, no.~6, pp. 1951--1964, 2022.

\bibitem{keskin2021jrc}
M.~F. Keskin \emph{et~al.}, ``{MIMO-OFDM} joint radar-communications: Is {ICI} friend or foe?'' \emph{IEEE Journal of Selected Topics in Signal Processing}, vol.~15, no.~6, pp. 1393--1408, 2021.

\bibitem{zhang2019jcrs}
J.~A. Zhang \emph{et~al.}, ``Multibeam for joint communication and radar sensing using steerable analog antenna arrays,'' \emph{IEEE Transactions on Vehicular Technology}, vol.~68, no.~1, pp. 671--685, 2019.

\bibitem{wukai2022isac}
K.~Wu \emph{et~al.}, ``Integrating low-complexity and flexible sensing into communication systems,'' \emph{IEEE Journal on Selected Areas in Communications}, vol.~40, no.~6, pp. 1873--1889, 2022.

\bibitem{wu2023dftsofdm}
Y.~Wu \emph{et~al.}, ``Sensing integrated {DFT}-spread {OFDM} waveform and deep learning-powered receiver design for terahertz integrated sensing and communication systems,'' \emph{IEEE Transactions on Communications}, vol.~71, no.~1, pp. 595--610, 2023.

\bibitem{lorenzo2020otfs}
L.~Gaudio \emph{et~al.}, ``On the effectiveness of {OTFS} for joint radar parameter estimation and communication,'' \emph{IEEE Transactions on Wireless Communications}, vol.~19, no.~9, pp. 5951--5965, 2020.

\bibitem{dehkordi2023otfs}
S.~K. Dehkordi \emph{et~al.}, ``Beam-space {MIMO} radar for joint communication and sensing with {OTFS} modulation,'' \emph{IEEE Transactions on Wireless Communications}, pp. 1--1, 2023.

\bibitem{wukai2023otfs}
K.~Wu \emph{et~al.}, ``{OTFS}-based joint communication and sensing for future industrial {IoT},'' \emph{IEEE Internet of Things Journal}, vol.~10, no.~3, pp. 1973--1989, 2023.

\bibitem{wu2023dftsotfs}
Y.~Wu \emph{et~al.}, ``{DFT}-spread orthogonal time frequency space system with superimposed pilots for terahertz integrated sensing and communication,'' \emph{IEEE Transactions on Wireless Communications}, 2023.

\bibitem{yuan2018beamforming}
H.~\vspace{0mm}Yuan \emph{et~al.}, ``Hybrid beamforming for mimo-ofdm terahertz wireless systems over frequency selective channels,'' in \emph{Proc. of IEEE Global Communications Conference (GLOBECOM)}, 2018.

\bibitem{yuan2020beamforming}
H.~Yuan \emph{et~al.}, ``Hybrid beamforming for terahertz multi-carrier systems over frequency selective fading,'' \emph{IEEE Transactions on Communications}, vol.~68, no.~10, pp. 6186--6199, 2020.

\bibitem{yan2022beamforming}
L.~Yan \emph{et~al.}, ``Energy-efficient dynamic-subarray with fixed true-time-delay design for terahertz wideband hybrid beamforming,'' \emph{IEEE Journal on Selected Areas in Communications}, vol.~40, no.~10, pp. 2840--2854, 2022.

\bibitem{gao2021beamforming}
F.~Gao \emph{et~al.}, ``Wideband beamforming for hybrid massive {MIMO} terahertz communications,'' \emph{IEEE Journal on Selected Areas in Communications}, vol.~39, no.~6, pp. 1725--1740, 2021.

\bibitem{dovelos2021mimo}
K.~Dovelos \emph{et~al.}, ``Channel estimation and hybrid combining for wideband terahertz massive mimo systems,'' \emph{IEEE Journal on Selected Areas in Communications}, vol.~39, no.~6, pp. 1604--1620, 2021.

\bibitem{zhai2021ofdma}
B.~Zhai \emph{et~al.}, ``{SS-OFDMA}: Spatial-spread orthogonal frequency division multiple access for terahertz networks,'' \emph{IEEE Journal on Selected Areas in Communications}, vol.~39, no.~6, pp. 1678--1692, 2021.

\bibitem{samara2023ofdm}
L.~Samara \emph{et~al.}, ``Adapt and aggregate: Adaptive {OFDM} numerology and carrier aggregation for high data rate terahertz communications,'' \emph{IEEE Journal of Selected Topics in Signal Processing}, 2023.

\bibitem{luo2019jcrs}
Y.~Luo \emph{et~al.}, ``Optimization and quantization of multibeam beamforming vector for joint communication and radio sensing,'' \emph{IEEE Transactions on Communications}, vol.~67, no.~9, pp. 6468--6482, 2019.

\bibitem{cheng2021dfrc}
Z.~Cheng \emph{et~al.}, ``Hybrid beamforming design for {OFDM} dual-function radar-communication system,'' \emph{IEEE Journal of Selected Topics in Signal Processing}, vol.~15, no.~6, pp. 1455--1467, 2021.

\bibitem{wang2022isac}
X.~Wang \emph{et~al.}, ``Partially-connected hybrid beamforming design for integrated sensing and communication systems,'' \emph{IEEE Transactions on Communications}, vol.~70, no.~10, pp. 6648--6660, 2022.

\bibitem{liu2019jsc}
F.~Liu and C.~Masouros, ``Hybrid beamforming with sub-arrayed {MIMO} radar: Enabling joint sensing and communication at {mmWave} band,'' in \emph{Proc. of IEEE International Conference on Acoustics, Speech and Signal Processing (ICASSP)}, 2019.

\bibitem{yan2020DAoSA}
L.~Yan \emph{et~al.}, ``A dynamic array-of-subarrays architecture and hybrid precoding algorithms for terahertz wireless communications,'' \emph{IEEE Journal on Selected Areas in Communications}, vol.~38, no.~9, pp. 2041--2056, 2020.

\bibitem{Elbir2021jrc}
A.~M. Elbir \emph{et~al.}, ``Terahertz-band joint ultra-massive {MIMO} radar-communications: Model-based and model-free hybrid beamforming,'' \emph{IEEE Journal of Selected Topics in Signal Processing}, vol.~15, no.~6, pp. 1468--1483, 2021.

\bibitem{yu2016altmin}
X.~Yu \emph{et~al.}, ``Alternating minimization algorithms for hybrid precoding in millimeter wave {MIMO} systems,'' \emph{IEEE Journal of Selected Topics in Signal Processing}, vol.~10, no.~3, pp. 485--500, 2016.

\bibitem{chen2022mimo}
Y.~Chen \emph{et~al.}, ``Millidegree-level direction-of-arrival estimation and tracking for terahertz ultra-massive mimo systems,'' \emph{IEEE Transactions on Wireless Communications}, vol.~21, no.~2, pp. 869--883, 2022.

\bibitem{levanen2021waveform}
T.~Levanen \emph{et~al.}, ``Mobile communications beyond 52.6 {GHz}: Waveforms, numerology, and phase noise challenge,'' \emph{IEEE Wireless Communications}, vol.~28, no.~1, pp. 128--135, 2021.

\bibitem{rikkinen2020THz}
K.~Rikkinen \emph{et~al.}, ``{THz} radio communication: Link budget analysis toward {6G},'' \emph{IEEE Communications Magazine}, vol.~58, no.~11, pp. 22--27, 2020.

\end{thebibliography}

\end{document}